\documentclass[twocolumn]{nictatechreport} 

\newcommand{\theTitle}{Rate Control Adaptation for Heterogeneous Handovers}
\newcommand{\theAuthor}{Olivier Mehani, Roksana Boreli, Guillaume Jourjon and Thierry Ersnt}
\newcommand{\theKeywords}{TFRC, congestion control, vertical handover, cross-layer, DCCP, transport protocol}

\usepackage[pdftex, unicode, pdfstartview=FitV,
  colorlinks=true, linkcolor=black, citecolor=black, urlcolor=blue,
  pdftitle={\theTitle}, pdfauthor={\theAuthor}, pdfkeywords={\theKeywords}]{hyperref}
\usepackage[utf8]{inputenc}
\usepackage[T1]{fontenc} \usepackage{aeguill}

\usepackage{amssymb}

\usepackage{graphicx} 
\usepackage{mathtools}
\usepackage{booktabs}
\usepackage{multirow}
\usepackage{slashbox}
\usepackage{subfig}
\usepackage{textcomp}
\usepackage{colortbl}
\usepackage{xcolor}
\usepackage[style=ieee,backend=bibtex]{biblatex} 
\usepackage{verbatim} 

\def\texturL#1{\texttt{\url{#1}}\endgroup}

\newcommand{\latinlocution}[1]{\textit{#1}}
\newcommand{\eg}{\latinlocution{e.g.}}
\newcommand{\ie}{\latinlocution{i.e.}}
\newcommand{\perse}{\latinlocution{per se}}

\newcommand{\sysfun}[1]{\texttt{#1}}
\newcommand{\sysvar}[1]{\texttt{#1}}
\newcommand{\rwnd}{\sysvar{rwnd}}
\newcommand{\cwnd}{\sysvar{cwnd}}

\newcommand{\ns}{\textit{ns}}

\begin{document}


\title{\theTitle} 
 


\author{Olivier Mehani,\textsuperscript{1,$\star$} Roksana Boreli\textsuperscript{1,2} 
Guillaume Jourjon,\textsuperscript{1} and Thierry Ernst\textsuperscript{3,4}}  
\affiliation{$^\star$Corresponding author: \email{olivier.mehani@nicta.com.au}\\ 
\textsuperscript1Nicta, Sydney. Eveleigh, NSW, Australia, \email{first.last@nicta.com.au} \\ 
\textsuperscript2University of New South Wales, Sydney, Australia \\ 
\textsuperscript3Mines ParisTech, Paris, France \\ 
\textsuperscript4Inria, Rocquencourt, Paris, France, \email{thierry.ernst@inria.fr}} 

\reportnumber{6163} 


\frontmatter 
\begin{abstract}

We present enhancements to the TCP-Friendly Rate Control mechanism (TFRC)
designed to better handle the intermittent connectivity occurring in mobility
situations. Our aim is to quickly adapt to new network conditions and better
support real-time applications for which the user-perceived quality depends on
the immediate transmission rate. We propose to suspend the transmission before
disconnections occur, in a way inspired by Freeze-TCP, and extend the solution by
probing the network after reconnecting to enable full use of the newly available
capacity.

We first introduce a numerical model of TFRC's performance after a network
handover and use it to evaluate the potential performance gains for realistic
network parameters.  We then describe a set of additions to TFRC to achieve
these gains.  Implementations within the Datagram Congestion Control Protocol
(DCCP) for \ns-2 and Linux have been adapted to support these enhancements.
Comparisons of experimental results for the original and modified DCCP are
presented for a number of example mobility scenarios.

We thus show how the proposed modifications enable faster recovery after
disconnected periods as well as significantly improved adjustments to the newly
available network conditions and an improvement in the quality of experience
(QoE) for video-streaming applications.

\end{abstract}



\mainmatter 

\section{Introduction}
\label{sec:intro}

In recent years, there has been a shift towards increased use of mobile devices
for Internet access and the use of real-time applications such as multimedia
streaming, VoIP and video conferencing in mobile environments.  This has been in
line with the increased capacity and widespread coverage of wireless
technologies, primarily cellular mobile (3--4G) and Wi-Fi. The vast majority of
today's mobile devices, in fact, include both 3G and Wi-Fi access options.
Mobility support for moving between either the same technology (horizontal
handovers) or cross-technology networks (vertical handovers) is supported by
technologies such as Mobile IP~\cite{rfc6275}.  However, the supporting
transport protocols used for the emerging applications are still those
which carried traffic for wired devices.

\begin{figure}[hb] 
  \centering 
  \resizebox{.85\columnwidth}{!}{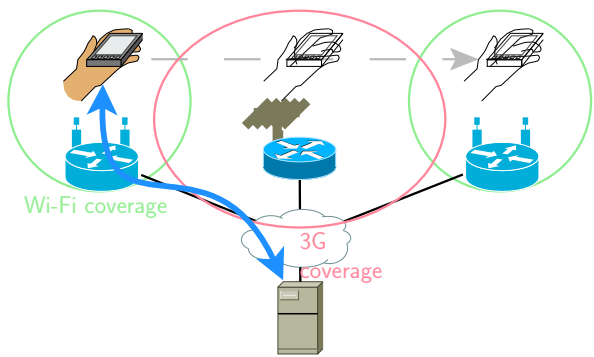}
  
  \caption[Generic use-case scenario]{Generic use-case scenario: vertical hand
  overs between a number of access networks with different characteristics.}

  \label{fig:heterogeneous}
\end{figure}

In this work, we consider \emph{break-before-make} (also sometimes called
\emph{hard}) handovers between networks with heterogeneous characteristics. In
our generic scenario,
depicted in \figurename~\ref{fig:heterogeneous}, 
a wireless mobile node (MN) moves between two or more networks while having
established sessions with fixed correspondent nodes (CN) in the Internet. The
handover can be between two access points of the same technology, that is,
\emph{horizontal}, or \emph{vertical}, between heterogeneous wireless networks
such as 3G to Wi-Fi.  Even though network mobility schemes hide most of the
complexity of roaming from the upper layers, cross-technology handovers usually
result in short, predictable disconnections during which network packets are
bound to be lost. As most available congestion control mechanisms interpret such
losses as an indication of congestion, they do not adapt
well~\cite{2002tsaoussidis_tcp_mobility_open_issues}.

UDP is currently often used to carry real-time traffic. As it does not provide
congestion control, it is not subject to this type of problem. However, it is
argued in~\cite{1999floyd_end2end_congestion_control} that congestion control
mechanisms should always be used in a shared internet.  TCP is not a viable
option for this type of content as retransmitted packets may already have gone
past their usefulness deadline by the time they reach the receiver.  To bridge
this gap, the Datagram Congestion Control Protocol
(DCCP)~\cite{2006kohler_dccp,rfc4340} has been proposed as a non-reliable but
congestion-aware transport protocol. DCCP can make use of the TCP-Friendly Rate
Control mechanism
(TFRC)~\cite{2000floyd_equation-based_unicast,2003widmer_equation-based_unicast_multicast,rfc5348},
which replicates TCP's response to adjust to the network capacity.  As it mimics
TCP's congestion control response, TFRC experiences similar
problems~\cite{2004gurtov_tfrc_vertical_handovers}.

In this article, we study the behaviour of TFRC over heterogeneous handovers and
propose mobility-aware enhancements.  We first analytically model the effects of
handovers on TFRC in order to quantitatively derive the potential for
improvement. This model details a complete handover and allows to evaluate the
number of lost packets during the disconnection to estimate the under-usage of
the newly available network capacity after the reconnection. It is validated
through \ns-2 simulations.  We then present an end-to-end solution to better
cope with such events.  This solution results in the extension of the TFRC
protocol which we implement and evaluate both in \ns-2 simulations, and
experiments using our Linux code. 

Some other solutions to the handover problem have been proposed in the
literature. However, to the best of our knowledge, no fully integrated solution
handles changes in the path characteristics after a ``lossy'' handover.  Our
disconnection-tolerant modification of TFRC shares similarities in concept with
Freeze-TCP~\cite{2000goff_freezetcp} but has different target applications and
introduces further enhancements.  This solution also relies on explicit handover
notifications (in a way similar
to~\cite{2006montavont_mipv6_link-network_interaction} on link triggers) to be
informed of mobility events, but the end-to-end path characteristics are then
discovered uniquely at the transport layer.

The main benefits of our proposal include preventing unnecessary rate reductions
by the transport protocol, both by avoiding false congestion detection and
adapting faster to the newly available capacity after vertical or horizontal
handovers. The proposal also enables either side of the connection to suspend
traffic entirely for the duration of the handovers. Thus, the resulting
Freeze-DCCP/TFRC is a disconnection-tolerant congestion-controlled protocol for
datagrams. The use of standard header fields and options makes our proposal
backward compatible with existing implementations of DCCP. 

Our Linux implementation allows us to show that this solution is well suited for
real-time traffic while coping with heterogeneous mobile handovers with
predictable disconnections.  In particular, we focus on the quality of
experience (QoE)~\cite{2008kilkki_qoe} in the form of PSNR~\cite{2001ansi_psnr}
for video streaming. This performance evaluation has been done over an
OMF-enabled testbed~\cite{2010rakotoarivelo_omf}, allowing for reproducibility
and verification of the experiments.

This article is structured as follows: Section~\ref{sec:related} presents some
related work and background information.  Section~\ref{freezetfrc:disconnection}
uses simulations to highlight the impact of mobility-induced disconnections on
TFRC, then introduces and validates a numerical model of its behaviour, allowing
to estimate possible performance gains. Section~\ref{freezetfrc:freezing}
presents the proposed TFRC protocol modifications and their implementation into
Freeze-DCCP/TFRC; \ns-2 simulations and experimental results with a Linux
implementation are presented in section~\ref{freezetfrc:performance}.  Finally,
in Section~\ref{freezetfrc:conclusion}, we summarise this work and present
future research directions.

\section{Related Work}
\label{sec:related}

The research literature contains a large number of contributions to deal with
node mobility, at all the layers of the TCP/IP
stack~\cite{2007nazir_mobility-enabled_stack}, as well as discussion about their
validity~\cite{2004eddy_layer_mobility}. At the transport layer, the proposals
address the problem of sessions surviving the change of address of the
endpoints. Though some generic work address wireless issues, only a few works
focus on adaptation to inherently heterogeneous paths in vertical mobility
handovers.  Here, we remind the reader to the state of the art of
congestion control mechanisms for the Internet. We also summarise the problems
which arise in mobile and wireless environments, and review the proposed
mitigation techniques.

\subsection{Congestion-Controlled Transport Protocols}

It is important to fairly share the
resource. It is therefore recommended to always use congestion-controlled
transport protocols~\cite{1999floyd_end2end_congestion_control}. TCP is the
epitome of such mechanisms. Its standard mechanism~\cite{rfc5681,rfc3782}
(comprising a slow-start and a congestion avoidance phase based on AIMD) is most
notably based on a \emph{congestion window} (\cwnd) tracking the number of bytes
that can be in flight and a \emph{receiver window} (\rwnd) manipulated by the
receiver to provide flow control. TCP's congestion control is currently the
standard of fair sharing on the
Internet~\cite{1999floyd_end2end_congestion_control,rfc5166}.  There is however
some contention with respect to how this fairness should be
assessed~\cite{2007briscoe_fairness}.

The Stream Control Transmission Protocol (SCTP)~\cite{rfc4960} has been designed
to offer a more flexible choice of feature combinations than the all-or-nothing
dichotomy of TCP and UDP. It implements an AIMD-based congestion control
similarly to TCP, but provides transport for datagrams rather than byte streams.
One of its salient features is its capability to register multiple address for
the session's endpoints to provide a failover mechanism if the primary path were
to fail.  It has been leveraged for handovers
management~\cite{2008han_sctpfx,2009han_sctpmx}.

The Datagram Congestion Control Protocol (DCCP)~\cite{2006kohler_dccp,rfc4340}
has been proposed to provide congestion-controlled datagrams streams.  However,
contrary to SCTP, it does not enforce delivery reliability. This makes it
a well suited transport for application such as real-time multimedia streaming,
where timeliness of data arrival is more important than reliability.  DCCP has
been designed with modularity in mind, and several congestion control mechanisms
(identified by their \emph{Congestion Control Identifier}, CCID) can be chosen.
A TCP-like congestion control algorithm, CCID~2~\cite{rfc4341}, follows TCP's
\cwnd-based mechanism made of slow-start and AIMD.  CCID~3~\cite{rfc4342} uses
TFRC (see below). CCID~4~\cite{rfc5622} has also been proposed as a variant of
CCID~3 for small packets such as used for VoIP.

The TCP-Friendly Rate Control (TFRC) is not a transport protocol \perse, but an
equation-based rate control
mechanism~\cite{2000floyd_equation-based_unicast,2003widmer_equation-based_unicast_multicast,rfc5348}.
It uses network-gathered metrics like RTTs, number of lost packets (more
precisely, the \emph{loss event rate}) and the data rate observed by the
receiver to compute the allowed sending rate, following a model equation of the
throughput of TCP\ Reno 
under the same
conditions~\cite{1998padhye_tcp_model}. Its operation is described in more
details in Section~\ref{freezetfrc:standard}. The use of an equation-based rate
control makes the rate changes smoother, which is more appropriate for the
streaming of multimedia content than the abrupt changes introduced by
AIMD~\cite{rfc5348}. It has also been argued
in~\cite{2004chen_rate-based_ad-hoc} that such class of rate controls tends to
be more resilient to wireless losses.

\subsection{Wireless Issues and Proposed Solutions}

In a well maintained wired-only network, the \emph{only} possible source of
losses is a router dropping packets due to its queue being full, that is, a
congestion. In this context, packet losses can clearly be considered to be fully
equivalent to congestion events, as the aforementioned congestion control
mechanisms assume.  Wireless links, however, can experience losses for other
reasons such as those related to propagation impairments or collisions at the receiver.

TCP's AIMD does not handle such losses well, as it interprets them as signs of
congestion.  New update laws for \cwnd\ have therefore been proposed to overcome
this problem. TCP Westwood~\cite{2001mascolo_westwood} and
Westwood+~\cite{2004grieco_westwood-plus_comparison} estimate the end-to-end
capacity  based on the rate of acknowledgements (ACKs) and adapt the \cwnd\ to
match this estimate.

It is noted in~\cite{1997balakrishnan_comparison_improvements_tcp_wireless} that
multiple layer-2 mitigating solutions have also been proposed. Some standardised
MAC protocols implement mechanisms to remediate or even avoid these losses.  For
example \emph{Request to send/Clear to
send} (RTS/CTS) mechanisms can be used to reserve the channel between both
nodes and mitigate the hidden node problem.  Most MAC mechanisms also adjust the
physical data rate depending on the channel conditions to ensure the majority of
the packets can be successfully received.

These MAC techniques are however not entirely transparent to the upper layers
and, if they successfully recover from a link-layer loss, it is at the price of
an increased delay to transmit the packet or an overall rate reduction. Several
studies have confirmed the performance degradation of TCP on these wireless
media~\cite{2001xylomenos_tcp_issues_wireless,2003pilosof_tcp_fairness_wlan,2004benekos_tcp_measurements_802.11b,2005franceschinis_tcp_wifi_measurement}.

To alleviate the impact of serious degradations of the wireless channel on the
TCP congestion control, an enhancement named
Freeze-TCP~\cite{2000goff_freezetcp} attempts to detect such situations at the
receiver and to suspend temporarily (``freeze'') the sender. To this end, it
leverages the window-based flow control mechanism of TCP by advertising a null
\rwnd\ (\emph{zero-window advertisement}) when the wireless channel fades away.
When the channel is restored to a usable level, the receiver sends a non
zero-window advertisement, therefore resuming the sender's operation with the
same congestion window.  It was extended to address predictable fadings in
vehicular networks~\cite{2006baig_freezetcp_on-board_performance} or mitigate
the impact of case of vertical handovers~\cite{2009park_freezetcpv2}

As for TCP, losses due to contention in wireless LANs disrupt TFRC's rate
estimation~\cite{2008zhang_cross-layer_congestion-control_wlans}.
The authors of these works identify the specific case where TFRC's natural rate
increase during loss-less periods leads to the wireless medium being saturated.
The thus-created losses lead TFRC to reduce its rate, and eventually results in
an oscillating behaviour. The authors therefore suggest to limit the transport
protocol's sending rate control law to the data rate currently achievable by the
underlying wireless link. They further extend the proposal by adding a similar
constraint in order to fairly share the wireless link with other users.

\subsection{Adapatability to Changes of Network Characteristics}

Congestion control mechanisms usually rely on estimates of the network
characteristics aggregated over time. In vertical handovers, it is not unlikely
that the new network characteristics are very different from the previous one's.
It will however take some time for the internal estimates to converge to the new
values, and properly adapt the rate.

To address this problem, SCTP is extended,
in~\cite{2008shieh_sctp_vertical_handover}, with a packet-pair probing of the
failover path prior to switching traffic. Similarly,
MBTFRC~\cite{2006lin_mbtfrc} as well as the work
in~\cite{2006chen_link_capacity_tfrc_probe}, have been proposed to implicitely
identify changes in the network characteristics, and react to them adequately.
Packet-pair techniques are however questionned as to their ability to adequately
estimate the capacity of a loaded path~\cite{2001dovrolis_packet-pair}.
High-to-low capacity changes are addressed in
\cite{2001bansal_slowly_responsive_congestion_control} which proposes some
self-clocking \textit{à la} TCP to avoid overloading a slow or congested link. A
more aggressive extension is also proposed
in~\cite{2008li_improving_tfrc_bandwidth_handover}. To accomodate for
intermitent traffic without unduly limiting the rate, a faster restart has also
been proposed for TFRC~\cite{draft-ietf-dccp-tfrc-faster-restart-06}, which
probes the network more aggressively under the assumption that its condition
might not have changes much since the last estimate.

Our proposal differs from the approaches above in that it is an integrated
transport-level solution which addresses both high-to-low and low-to-high
handovers and adjusts the rate accordingly through in-band techniques. Moreover,
it also targets break-before-make events (of which \emph{make-before-breaks} are
a subset) and the resulting packet losses during the disconnected period.
Finally, it relies on explicit notifications (the only cross-layer information
needed), which allows for a more timely and accurate knowledge of path changes.

\section{Behaviour of TFRC During a Disconnection}
\label{freezetfrc:disconnection}

In this section, we investigate the issues TFRC faces when used in mobile
scenarios with heterogeneous handovers. We first describe the operation of TFRC
as standardised by~\cite{rfc5348}.
%
%
We then provide an example simulation highlighting some of the issues.
%
%
%
 The behaviour illustrated here  
is then modelled in order to quantify the
performance issues.

\subsection{Operation of Standard TFRC}
\label{freezetfrc:standard}

Based on feedback from the receiver, a standard TFRC sender controls its rate
$X$ following a model of TCP's throughput under the same
conditions~\cite{1998padhye_tcp_model},
\begin{gather}
  \label{eq:tfrc-ca}
  X_\mathrm{Bps} = T(p,R) = \frac{s}{R\sqrt{\frac{2p}3} + t_{RTO}\left( 3\sqrt{\frac{3p}8}
  \right)p\left( 1+32p^2 \right)}, \\
  \label{eq:tfrc-update-x}
  X \leftarrow \min(X_\mathrm{Bps},2X_\mathrm{recv})
\end{gather}
where $s$ is the packet size, $R$ the RTT, and $t_{RTO}$ the retransmit timeout
(usually $4R$, measured by the \verb#nofeedback# timer).  Parameters $p$ and
$X_\mathrm{recv}$ are reported by the receiver roughly every RTT and are,
respectively, the loss event rate and the current received rate. If no report
from the receiver is seen before $t_\mathrm{RTO}$ expires, the sender reduces
its allowed sending rate by halving the last used $X_\mathrm{recv}$.

As for TCP, a slow-start phase is also present to first adapt the rate
to the network path's capacity. During this phase, the sender updates its rate
once per RTT following
\begin{gather}
  X \leftarrow \min(2X, 2X_\mathrm{recv}),
  \label{eq:tfrc-ss}
\end{gather}
until the first loss is observed.  When the first loss occurs, the TFRC receiver
reports a loss event rate $p$ which reflects its observed throughput
$X_\mathrm{recv}$ before the loss. The value of $p$ is initialised by inverting
\eqref{eq:tfrc-ca}.\footnote{It is not specified in~\cite{rfc5348} how this
inversion should be done. Most implementations use a binary search, but
\cite{2007jourjon_tfrc_initialization} suggests a more CPU efficient method
based on a Newton search.}

When further losses are observed, the TFRC receiver computes $p$ as the
inverse of the weighted average of the lengths of the $n$ most recent loss
intervals $i_0,\ldots,i_{n-1}$. The length of a loss interval is measured in
number of packets successfully received. The average is computed using a vector
of decreasing weights $\vec w = [w_0,\ldots,w_{n-1}]$ as $S_0 =
\sum_{i=0}^{n-1}w_ii_i$.  The TFRC receiver actually keeps a history $\vec i =
[i_0,\ldots,i_{n}]$ of the last $n+1$ loss intervals.  This slightly larger
buffer is designed to avoid overly increasing the loss event rate when one of
these events has just happened. Indeed, when losses have just been experienced,
the size of the current loss interval $i_0$ starts increasing from 0. At first,
$i_0$ is so small that it would incorrectly drive $p$ up and needlessly reduce
the rate $X_\mathrm{Bps}=T(p,R)$. It is therefore ignored and the reported loss
event rate is still based on the previous $i_1,\ldots,i_n$ intervals. As these
values do not change anymore, $p$ is stationary during this period.  Taking $S_1
= \sum_{i=0}^{n-1}w_ii_{i+1}$, $p$ is computed in~\cite{rfc5348} as
\begin{gather}
  \label{eq:ImeanP}
  p = \frac1{i_\mathrm{mean}} = \frac{\sum_{i=0}^{n-1}w_i}{\max(S_0, S_1)}.
\end{gather}
Computing $p$ this way only considers the duration of loss-less periods and
is not related to that of those during which all packets are lost,
even if they span several RTTs.

As~\eqref{eq:tfrc-ca} has an inverse relation with the loss event rate $p$, TFRC
is not fit to work on networks with loss-inducing disconnections.  A temporary
break in the end-to-end path would indeed have several consequences.  First,
packets will needlessly use parts of the network path's capacity before being
dropped, resulting in losses. The sending rate will then gradually be reduced as
the \verb#nofeedback# timer expires. Upon reconnection, the transport protocol's
rate will therefore not match the network's characteristics and need some time
to re-adapt. We illustrate this behaviour in the next section.

Moreover, as the computation of the loss event rate is based on a history of
several loss events, TFRC reacts slowly to immediate decreases in $p$. In the
case of a cross-technology hand-off to an access networks with larger capacity,
it will therefore take a much longer time to adjust the rate to the newly
available capacity.

\subsection{Simulation of Mobile Handovers}
\label{sec:freezetfrc-mobile-handovers}

\begin{figure}[tb]
  \centering 
  \resizebox{.85\columnwidth}{!}{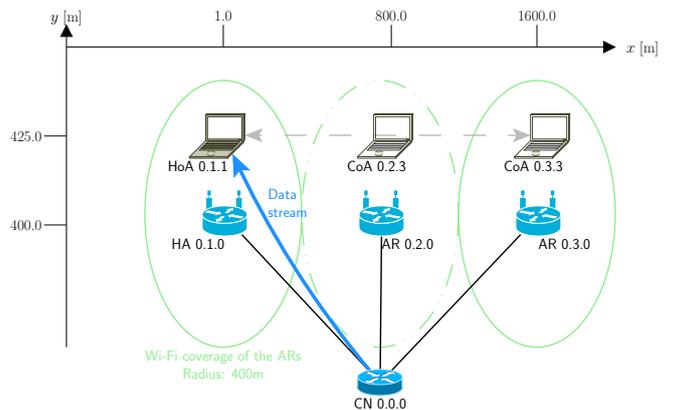}
  
  \caption[DCCP/TFRC mobility simulation scenario]{The basic simulation
  scenario. A mobile node (MN) moves back and forth between three adjacent (but
  not overlapping) wireless networks with different prefixes. The correspondent
  node (CN) sends a constant stream of data to the MN's home address (HoA) using
  DCCP/TFRC. For simplicity, the MN's home agent (HA) is set to be the access
  router (AR) of the first access network.  Addresses are expressed in \ns-2
  format.}

  \label{fig:systematic_simulation}
\end{figure}

To present an example of the adverse consequences of disconnections on the
sending rate of DCCP/TFRC in mobility situations, we ran several simulations
with \ns-2~\cite{2009vint_ns_manual}.  The simulation scenario consists of a
landscape of $800\times1600$\,m where a Mobile IPv6 (MIPv6, \cite{rfc3775})
mobile node (MN) moves between the coverage of three access routers (AR).
\figurename~\ref{fig:systematic_simulation} presents the simulated environment,
consisting of one backbone router and the three ARs providing adjacent but
non-overlapping wireless connectivity to the MN receiving traffic.  Simulations
were also run with the second base station disabled in order to observe the
behaviour of the data stream in the case of a more sporadically available
network coverage. 

\ns-2 was configured to simulate a regular single-rate 11\,Mbps 802.11b wireless
channel. Some parameters had to be adjusted to obtain the desired simulation
conditions: the wireless reception threshold has been fine-tuned to simulate a
400\,m Wi-Fi range. \tablename~\ref{tab:ns80211} summarises these configuration
changes.  We have also ported the MobiWan MIPv6 support~\cite{2001ernst_mobiwan}
and DCCP module~\cite{2004mattsson_dccp_ns2} to version 2.33 of this simulator,
and updated these to the latest versions (at the time) of their respective
specifications~\cite{rfc3775,rfc5348}.\footnote{These updated patch-sets are
available at \url{http://www.nicta.com.au/people/mehanio/nsmisc/}.} All kinds of
route optimisations for MIPv6 were disabled.

\begin{table}[tb]
  \centering

  \caption[Simulations parameters adjusted from \ns-2's defaults]{Parameters
  adjusted from \ns-2's defaults.}

  \label{tab:ns80211}
  \begin{tabular}{cc}
    \toprule
    \textbf{\ns-2 parameter} & \textbf{Value} \\
    \midrule
    \multicolumn{2}{c}{\textbf{11\,Mbps 802.11b channel}} \\
    \midrule
    \verb#Phy/WirelessPhy bandwidth_# & \verb#11Mb# \\
    \verb#Phy/WirelessPhy freq_# & \verb_2.472e9_ \\
    \verb#Mac/802_11 dataRate_# & \verb_11Mb_ \\
    \verb#Mac/802_11 basicRate_# & \verb_1Mb_ \\
    \midrule
    \multicolumn{2}{c}{\textbf{Miscellaneous}} \\
    \midrule
    \verb#Phy/WirelessPhy RXThresh_# & \verb_5.57346e-11_ \\
    \verb#Agent/MIPv6/MN bs_forwarding_# & \verb_0_ \\
    \verb#Agent/MIPv6/MN rt_opti_# & \verb_0_ \\
    \bottomrule
  \end{tabular}
\end{table}

\begin{figure}[tb]
  \centering
  \includegraphics[width=.75\columnwidth]{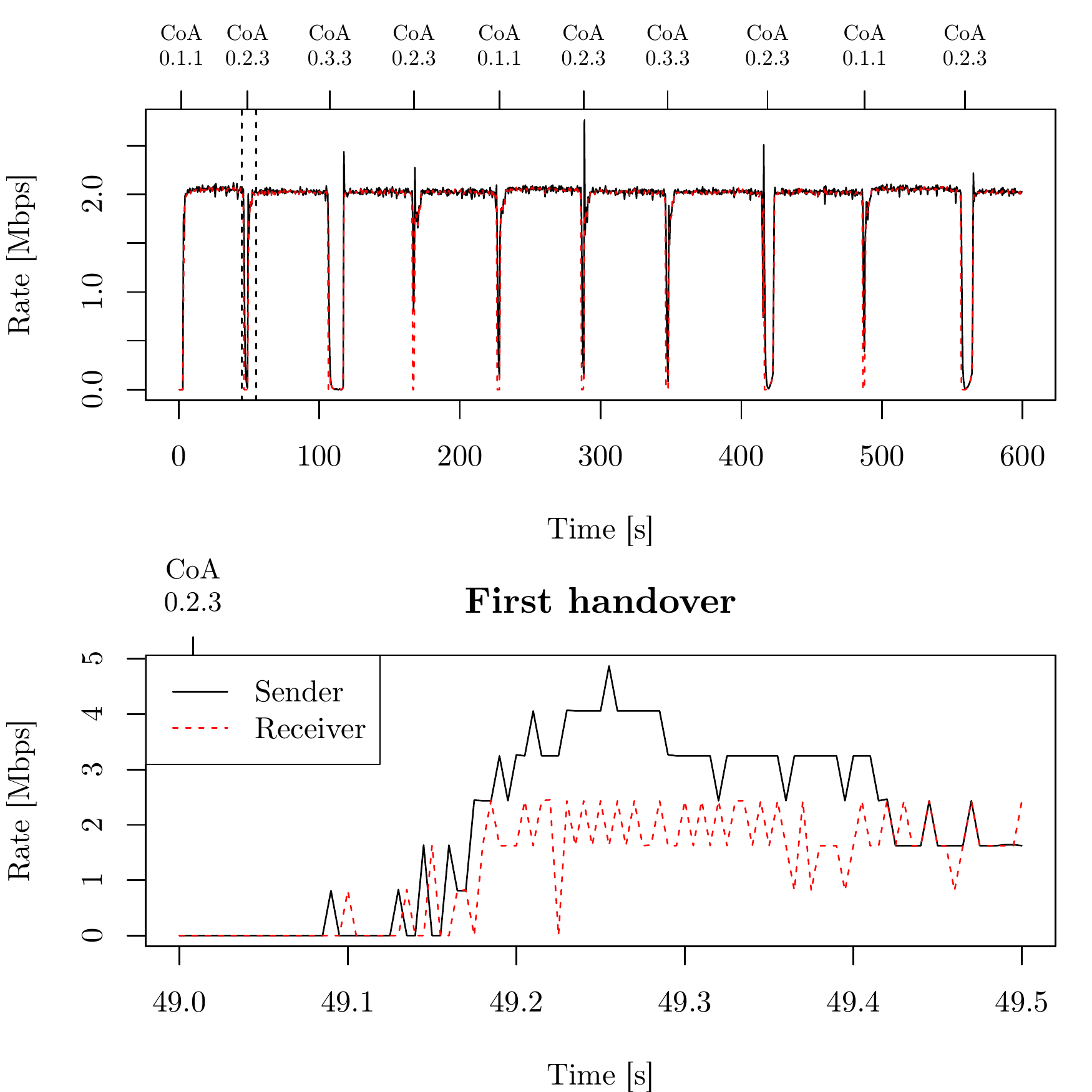}

  \caption[DCCP/TFRC data flow moving through adjacent Wi-Fi access
  networks]{DCCP/TFRC data flow moving through adjacent Wi-Fi access
  networks. Labels on the top axis represent the time when the new CoA has
  been configured and is fully usable.}
  
  \label{fig:sys-dccp-3bs}

\end{figure}

\begin{figure}[tb]
  \centering
  \includegraphics[width=.75\columnwidth]{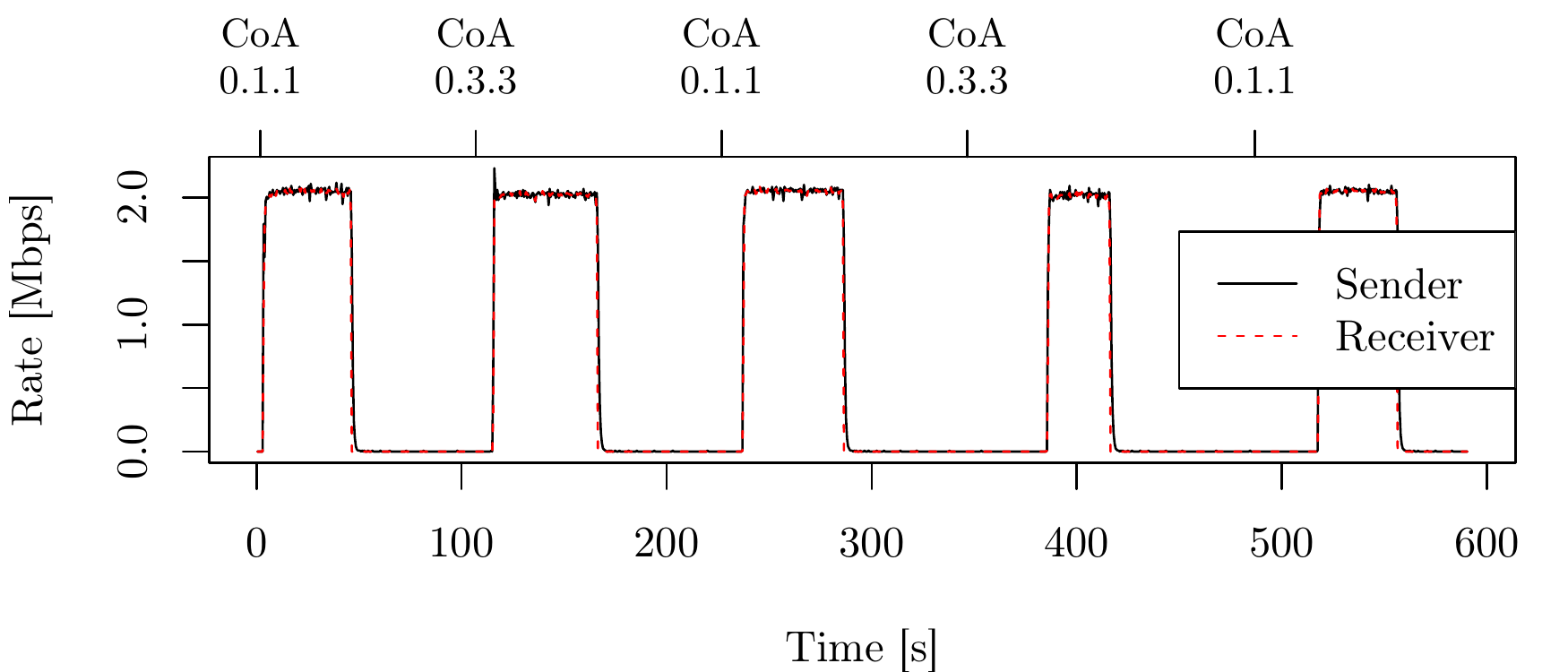}

  \caption[DCCP/TFRC data flow moving through non-adjacent Wi-Fi access
  networks]{DCCP/TFRC data flow moving through non-adjacent Wi-Fi access
  networks.}

  \label{fig:sys-dccp-2bs}
\end{figure}

In the first scenario, the MN moves back and forth at constant speed between all
three ARs (from adjacent Wi-Fi networks), losing connectivity with the current
one, associating with the new one, and re-establishing its mobility bindings
with its HA. \figurename~\ref{fig:sys-dccp-3bs} shows that, in addition to the
delay to associate with the new AR and re-establish the bindings when the
previous link breaks, there is a delay between the time when a care-of address
(CoA, the locator in the current access network) is fully configured on the new
access network, and when the rate of TFRC is reinstated: 100\,ms until it
restarts, but 500\,ms until it is fully restored.
\figurename~\ref{fig:sys-dccp-2bs} shows the results for a similar scenario
where the second access point has been disabled, thus creating a period of
complete lack of connectivity.  The previous delay effect becomes much larger,
up to 50\,s in this case.

In the next section, we model this behaviour in order to evaluate the performance
issues in terms of lost packets, delay until restart and ``wasted'' capacity.


\subsection{Numerical Model of TFRC's Behaviour}
\label{sec:model}

%
In order to confirm the generality of the example of the previous section, and
quantify the highlighted impact, we
%
%
%
introduce a model of TFRC's behaviour when a disconnection occurs. This model is
used to derive the number of packets that are lost during the disconnection, the
delay before TFRC resumes sending after a reconnection, the available capacity
and the time it takes to adapt to the new characteristics. After validating it
with \ns-2 simulations, we evaluate these metrics for various typical horizontal
and vertical handover scenarios. This shows that there is ample room for better
management of these events. A summary of the symbols used throughout this
section is given in \tablename~\ref{tab:notations}.

\begin{table}[tb]
  \centering

  \caption[Notations for TFRC modelling]{Notations used for the analysis of
  TFRC.}

  \label{tab:notations}
  \addtolength{\tabcolsep}{-1ex}
  \begin{tabular}{cp{7.5cm}}
    \toprule
    \textbf{Sym.} & \hfill\textbf{Meaning}\hfill\strut \\
    \midrule
    \multicolumn{2}{c}{\textbf{During disconnection}} \\
    \cmidrule{1-2}
    $t_\mathrm{RTO}^i$ & Duration of NFI $i$ \\
    $X^i$ & Sender rate during NFI $i$ ($X_d = X^0$) \\
    $i_x$ & First NFI so that $X^{i_x} = s/t_\mathrm{mbi}$ \\
    $i_t$ & First NFI so that $t_\mathrm{RTO}^{i_t}$ starts increasing \\
    $n_\mathrm{lost}$ & Number of packets lost during the disconnection \\
    \cmidrule{1-2}
    \multicolumn{2}{c}{\textbf{After reconnection}} \\
    \cmidrule{1-2}
    $t_\mathrm{idle}$ & Time before the first packet is sent after reconnection\\
    $n_R^\varepsilon$ & Number of RTTs on the new network before $R_r^i$ is within $\varepsilon$ of $R_\mathrm{new}$ \\
    $n_\mathrm{pkts}^i$ & Total number of packets sent after RTT $i$ on the new network \\
    $n_\mathrm{wasted}^*$ & Number of packets that could have been sent \\
    \bottomrule
  \end{tabular}
\end{table}

\subsubsection{During the Disconnection}

During the disconnection, the TFRC senders keeps transmitting packets. However,
no feedback from the receiver can be received, and the sending rate is
gradually reduced. Here, we evaluate this rate, and the number of packets which
get sent but cannot be received.

\begin{figure*}[tb]
  \subfloat[Internal parameters]{
    \resizebox{.32\textwidth}{!}{
      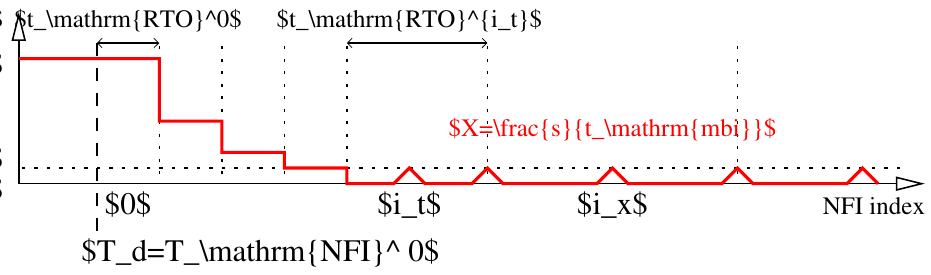
    }
    \label{fig:regular-tfrc-internal}
  }
  \subfloat[Rate behaviour]{
    \resizebox{.32\textwidth}{!}{
      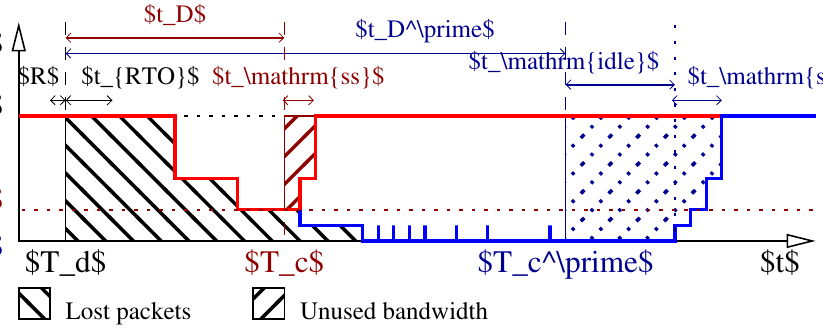
    }
    \label{fig:regular-tfrc}
  }
  \subfloat[Larger capacity]{
    \resizebox{.32\textwidth}{!}{
      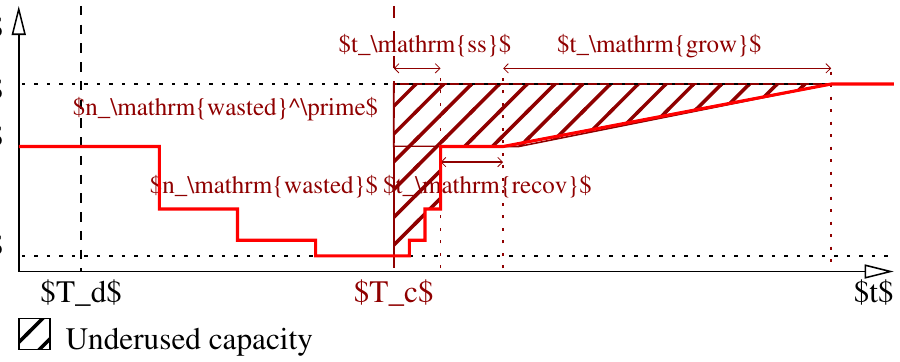
    }
    \label{fig:regular-tfrc-bigger}
  }

  \caption[TFRC sender's internal parameters and rate behaviours.]{TFRC sender's
  internal parameters and rate behaviours.
  \subref{fig:regular-tfrc-internal}~Evolution of internal parameters after a
  disconnection. \subref{fig:regular-tfrc}~Rate behaviour in a period with no
  connectivity. Two cases are shown, with different times of reconnection: at
  $T_c$, a period $t_D$ has elapsed which is short enough that TFRC's rate did
  not reach its minimum (red) and at $T_c^\prime$, when the time $t_D^\prime$
  elapsed since the disconnection was sufficient for $X$ to reduce to
  $s/t_\mathrm{mbi}$ (blue); an additional delay $t_\mathrm{idle}$ is present in
  the latter case before TFRC starts restoring its rate.
  \subref{fig:regular-tfrc-bigger}~After a reconnection, TFRC does not adapt
  quickly to higher capacities. It slowly uses more capacity as $p$ decreases.}

\end{figure*}

\paragraph{Evolution of Internal Parameters} We evaluate the changes in sender
rate $X$ during a disconnection, as well as the \verb#nofeedback# timer's
period, $t_\mathrm{RTO}$. Both values have an impact on the number of packets
lost during the disconnection and the rate recovery after the reconnection.
\figurename~\ref{fig:regular-tfrc-internal} represents the evolution of these
parameters.

Just before the disconnection occurs, at $T_d$, the sender sends at rate $X=X_d$.
In the following, this is assumed to be the nominal TFRC rate that the
underlying link can support. Consequently, the receiver measures and reports an
$X_\mathrm{recv}$ roughly equal to $X_d$.  Thus, \eqref{eq:tfrc-update-x}
is limited by $X_\mathrm{Bps}$.

In the absence of feedback, the TFRC sender halves its allowed sending rate
every time the \verb#nofeedback# timer expires by reducing its local estimate of
$X_\mathrm{recv}$. When $X$ becomes small, $t_\mathrm{RTO}$ is increased to
cover the transmission of at least two packets.

For convenience, we segment the disconnected period into \emph{no feedback
intervals} (NFI). An NFI is the interval between two consecutive expirations of
the \verb#nofeedback# timer.%
\footnote{An NFI is the same concept as the NFT of \cite{draft-ietf-dccp-tfrc-faster-restart-06}.}
 NFIs are indexed starting at
$i=0$. The first expiration of the \verb#nofeedback# timer marks the end of NFI
0. Hence, the effects of this timeout start at the beginning of NFI 1.  The rate
then gradually decreases until it reaches its minimum value during NFI $i_x$.

Every NFI, the sender halves the value of $X_\mathrm{recv}$, which then
drives~\eqref{eq:tfrc-update-x}. In the worst situation, $X$ can reduce to the
minimal value of one packet every 64 seconds ($s/t_\mathrm{mbi}$).  Taking $i_x$
as the NFI during which $2X_\mathrm{recv}^{i_x}$ drops below $s/t_\mathrm{mbi}$,
the sender rate can be expressed as
\begin{gather}
  X^i = \begin{cases}
    \frac{X_d}{2^{i}} & \text{if $0 \leq i < i_x$,} \\
    \frac{s}{t_\mathrm{mbi}} & \text{otherwise,}
  \end{cases}
  \label{eq:numerical-x-i} \text{ with } 
  i_x =  \left\lceil \log_2\frac{X_d\cdot t_\mathrm{mbi}}s \right\rceil,
\end{gather}
where $\lceil\cdot\rceil$ is the ceiling operator.

Additionally, the \verb#nofeedback# timer, initially set to
$t_\mathrm{RTO}^0=4R$, increases when the sending rate becomes smaller than
$2s/4R$. Assuming $X_d\ge2s/4R$ and taking $i_t$ as the NFI during which
$2s/X^{i_t}$ becomes larger than $4R$, the duration of NFI $i$ is then
\begin{gather}
  t_\mathrm{RTO}^i = \begin{cases}
    4R & \text{if $i < i_t$,} \\
    \frac{2s}{X^i} & \text{otherwise,}
  \end{cases}
  \label{eq:numerical-trto-i} \text{ with }
  i_t =  \left\lceil \log_2\frac{2R\cdot X_d}s \right\rceil.
\end{gather}
Note that~\eqref{eq:numerical-trto-i} is only valid for $R<t_\mathrm{mbi}/2$, in
which case $i_t \le i_x$. Otherwise, $4R$ is larger than the time to send 2
packets at the lowest rate, and $i_t$ is considered to be $+\infty$.

\paragraph{Packet Losses}

Though the number of losses happening during the single loss event of the
handover does not directly impact TFRC's sender rate, they are an unnecessary
burden on the rest of the network which could be better used for other traffic,
for which data can actually be delivered to the destination. It is interesting
to quantify this charge on the network in the form of the number of packets
which will be lost during the disconnected period.

\figurename~\ref{fig:regular-tfrc} shows the evolution of the sender rate during
a handover. Two cases are represented, for different reconnection times $T_c$
and $T_c^\prime$. They respectively occur before and after the sender's
estimation of the receiver rate has reduced to less than one packet per RTT.

Time $t_D = T_c - T_d$ is the length of the disconnected period.  All the
packets sent during this period are lost.  The number of lost packets when the
reconnection occurs, after $n_D$ NFI (such that
$\sum_{i=0}^{n_D}t_\mathrm{RTO}^i \ge t_D$), can be estimated using
\begin{gather}
  \label{eq:numerical-packet-losses}
  n_\mathrm{lost} = \begin{cases}
    \left\lfloor \frac{t_DX^0}s \right\rfloor & \text{if $t_D \le t_\mathrm{RTO}^0$,} \\
      \left\lfloor \frac{t_\mathrm{RTO}^0X^0}s + \sum_{i=1}^{i_D} \frac{t_\mathrm{RTO}^iX^i}s \right\rfloor & \text{otherwise,}\\
  \end{cases}
\end{gather}
where $i_D=n_D-1$ is the index of the $n_D^\mathrm{th}$ NFI and
$\lfloor\cdot\rfloor$ is the floor operator.

\subsubsection{After the Reconnection}
\label{sec:after-reco}

After having reconnected, packets and feedback messages can be exchanged again,
but some waiting time due to exponential backoff may delay this restart.  The
loss event rate $p$ gets updated on the first feedback message, then so is the
rate. When the rate can increase again, it will however reflect the parameters
observed on the previous network path, which may be inappropriate for the new
access network.

\paragraph{Variation of the Loss Event Rate}

The losses will only be noticed by the receiver after reconnecting.  In
\cite{rfc5348}, the authors specify that the expected arrival time of a lost
packet is interpolated using those of both packets received directly before and
after the loss. Multiple losses over the disconnected period will then be
considered part of the same loss event starting in the middle of the
disconnected period.

Following the procedure described in Section~\ref{freezetfrc:standard}, the
evolution of $p$ can be in three different phases depending on $S_0$ and $S_1$
(defined in that section):
\begin{description}
  \item[no loss] when $S_0\ge S_1$, $p$ gradually decreases as the number of
    received packets, in $i_0$, increases;
  \item[first loss observed] makes $S_0<S_1$ which stabilises $p$ until the current loss
    interval $i_0$ becomes large enough that the inequality is reversed;
  \item[new loss observed] when $S_0<S_1$, $\vec i$ gets shifted which increases
    $p$ as~\eqref{eq:ImeanP} now has a smaller denominator.
\end{description}

In the congestion avoidance state, the variation of loss event rate ($\Delta p$) can be in different
ranges depending which of the three phases the sender is in.
\begin{gather}
  \begin{cases}
    \Delta p = 0, \qquad \text{after the first loss has been observed,} \\
    \Delta p > 0, \qquad \text{when more losses are observed,} \\
    \Delta p_\mathrm{min}(\Delta n_\mathrm{pkts},p_\mathrm{prev}) \le \Delta p < 0, \qquad \text{otherwise (no loss).}
  \end{cases}
  \label{eq:numerical-p-variation}
\end{gather} 
The lower bound of $\Delta p=p-p_\mathrm{prev}$ when no losses occur can be derived
using~\eqref{eq:ImeanP} to estimate $p$. If all
$\Delta n_\mathrm{pkts}$ packets sent since the last feedback, which reported
$p_\mathrm{prev}$, have been received and taken into account in the receiver's
next feedback, the reduction in $p$ will be the lower-bound
\begin{gather}
  \Delta p_\mathrm{min}(\Delta n_\mathrm{pkts},p_\mathrm{prev}) = 
  \underbrace{\frac{\sum_{i=0}^{n-1}w_i}{w_0\Delta n_\mathrm{pkts} +
  (\sum_{i=0}^{n-1}w_i)/p_\mathrm{prev}}}_{p=1/i_\mathrm{mean}}-p_\mathrm{prev}.
  \label{eq:numerical-delta-p}
\end{gather}

\paragraph{Event Timing}

While the time base of changes of the disconnected sender are regulated by the
length of its retransmit timeout, feedback from the receiver takes on this role
once reconnected. The receiver sends periodic feedback messages at least once
per RTT. For the rest of this section, time will then be segmented in units of
RTTs. Indices $i$ now refer to how many of those have elapsed since the
reconnection, rather than NFI as previously.

The periodicity of sender-side events triggered by these feedback messages will
follow the RTT of the visited network.  While this value is mostly
stationary for a given network, the sender does not use it directly for its
computations, most notably that of the sending rate. To ensure a smooth
evolution, it uses an exponentially weighted moving average of the past samples
to estimate the RTT. After feedback message $i$ allowing to sample RTT
$R_\mathrm{new}$ of the new network, the sender's estimate of the RTT is
  $R_r^i = q R_r^{i-1} + (1-q) R_\mathrm{new},$
with $0<q<1$ ($q=0.9$ in~\cite{rfc5348}).  Just after the reconnection, before
the first feedback message has been received, the estimate completely reflects
the RTT of the previous network, $R_r^0 = R_\mathrm{old}$.
Expanding the series, this estimate can be expressed as
\begin{align}
  \label{eq:Rri}
  R_r^i &= (1-q)R_\mathrm{new}\underbrace{\sum_{j=0}^{i-1}q^j}_{\frac{q^0 - q^i}{1-q}} + q^i R_r^0 
  = \left(1-q^i\right)R_\mathrm{new} + q^iR_\mathrm{old}.
\end{align}
It will then evolve from a value representing of the RTT on the previous
network, $R_\mathrm{old}$, to that of the new network. Depending on the
difference ratio between these two values, a variable number of samples will be
needed for the sender to have an accurate estimate of $R_\mathrm{new}$. The
number of samples, denoted $n_R^\varepsilon$, needed to have an estimate within
$\varepsilon$ of the actual value, that is,  $\left| R_r^{n_R^\varepsilon} -
R_\mathrm{new} \right| \le \varepsilon$, is
\begin{gather}
  \label{eq:iRvarepsilon}
  n_R^\varepsilon = \left\lceil\frac{\ln\varepsilon - \ln\left| R_\mathrm{old}-R_\mathrm{new} \right|}{\ln q}\right\rceil.
\end{gather}
Assuming that the order of magnitude can vary from the millisecond to the second
depending on the network technology and load, it can take up to almost 30 RTTs
on the new network for the estimate to be accurate within a 5\,\% margin.

It is important to note that the TFRC equation~\eqref{eq:tfrc-ca} is directly
dependent on this estimate. Given a static $p=p_r$ after the reconnection, as
per~\eqref{eq:numerical-p-variation}, successive samples of the new RTT will
refine the estimate which will in turn impact $X_\mathrm{Bps}$. As
$t_\mathrm{RTO}=4R$, $X_\mathrm{Bps}$ can be expressed as a function of any of
its previous values $T(p_r, R^\prime)$ and the associated RTT estimate $R^\prime$
\begin{align} 
  \label{eq:XiBps}
  X_\mathrm{Bps}^i &= T(p_r, R_r^i) = \frac{R^\prime}{R_r^i} T(p_r, R^\prime).
\end{align}
The most useful such relation involves $X_d=T(p_r,R_\mathrm{old})$:
$X_\mathrm{Bps}^i = (R_\mathrm{old}/R_r^i)X_d$; upon reconnection, the new
TFRC rate, as compared to that before the disconnection, is thus only
dependent on the ratio of the current and previous estimations of the RTT, as
$p_r$ is computed from $S_1$.

\paragraph{Number of Sent Packets}

When the connection is re-established and the TFRC sender restarts sending
packets, it goes through a few phases before being able to resume a rate
appropriate to the current network. Depending on the differences between the
previous and the current networks' characteristics, the duration (or existence)
of these phases will vary. An important factor impacting theses phases is the
number of packets that have been sent since the reconnection. In the following,
it will be expressed as
  $n_\mathrm{pkts}^i = 1/s\sum_{j=0}^i R_r^jX_r^j$,
where $i$ is the RTT at the end of which the packets are counted and $X_r^i$
is the sending rate during that RTT, as detailed below.

\paragraph{Unused Capacity}

When connectivity is re-established, two factors can prevent the TFRC sender
from fully utilising the available capacity instantaneously.  First, when
feedback is received, the sending rate is not resumed directly, but through a
phase similar to slow-start (\eg, at $T_c$ in
\figurename~\ref{fig:regular-tfrc}).  Second, the sender is not directly aware
of the reappearance of connectivity. It has to wait for a packet to be
acknowledged by the receiver. As the sending rate has been gradually reduced,
said packet may not be immediately sent (\eg, $t_\mathrm{idle}$ after
$T^\prime_c$ in \figurename~\ref{fig:regular-tfrc}). 

If the sending rate when reconnecting, $X_c=X^{n_D}$, is small, the delay
$s/X_c$ between the transmission of two subsequent packets becomes significant.
When connectivity is recovered, it can take up to this delay before the first
packet is sent. The average idle time after reconnecting can be expressed as
$t_\mathrm{idle} = s/2X_c$.

After this delay, the sender eventually starts increasing the rate. Packets are
first sent at rate $X_c$. Every RTT, a feedback is received with the current
value of $X_\mathrm{recv}$. According to~\eqref{eq:tfrc-update-x}, the rate can
then be updated to twice this value until $X_r^i$ reaches the rate allowed by
the TFRC equation, $X_\mathrm{Bps}^i = (R_\mathrm{old}/R_r^i)X_d$. During
the slow-start RTT $i$, the sender rate is
\begin{gather}
  \label{eq:numerical-xrss}
  X_r^i = 2^i X_c.
\end{gather}
Assuming the new network provides the same capacity as the previous one, the
average number of packets that could additionally be sent is 
\begin{gather}
  n_\mathrm{wasted} = \frac1s\left(t_\mathrm{idle} \cdot X_d + \sum_{i=0}^{n_\mathrm{ss}} R_\mathrm{new}
  \left( X_d - X_r^i \right)\right).
  \label{eq:numerical-packets-wasted}
\end{gather}
Parameter $n_\mathrm{ss}$ is such that $X_r^{n_\mathrm{ss}} \ge
X_\mathrm{Bps}^{n_\mathrm{ss}}$. The development of this inequality
using~\eqref{eq:Rri}, \eqref{eq:XiBps} and~\eqref{eq:numerical-xrss} leads to
\begin{gather}
  \frac{R_\mathrm{new}}{R_\mathrm{old}}2^{n_{ss}} + \left(1-\frac{R_\mathrm{new}}{R_\mathrm{old}}\right)(2q)^{n_{ss}} > \frac{X_d}{X_c},
  \label{eq:numerical-nss}
\end{gather}
which is not linear in $n_\mathrm{ss}$. It thus cannot be solved in a purely
analytical fashion. In our implementation of the model, we used the
Newton-Raphson method%
%
~\cite{1995ypma_newton-raphson} with 
\begin{gather}
  f(n_{ss}) = \frac{R_\mathrm{new}}{R_\mathrm{old}}2^{n_{ss}} + \left(1-\frac{R_\mathrm{new}}{R_\mathrm{old}}\right)(2q)^{n_{ss}} - \frac{X_d}{X_c}\quad\text{and} \\
  f^\prime(n_{ss}) = \frac{R_\mathrm{new}}{R_\mathrm{old}}2^{n_{ss}}\ln2 + \left(1-\frac{R_\mathrm{new}}{R_\mathrm{old}}\right)(2q)^{n_{ss}}\ln2q.
\end{gather}
This allowed us to solve~\eqref{eq:numerical-nss} for $n_{ss}$ in only a few
iterations starting from an arbitrary $n_{ss0} = 10$, regardless of the network
parameters.

\paragraph{Networks with Larger Capacity} The estimate of the loss event rate
$p$ is designed to evolve smoothly. This may cause an additional under-usage of
the available capacity in case the MN connects to a network with a higher
capacity, $X_\mathrm{max}$, than the previous link.  Depending on the difference
in capacity from said previous network, it may take an unacceptably long time
for the sender to eventually occupy the full available capacity.
\figurename~\ref{fig:regular-tfrc-bigger} shows how TFRC slowly adapts to the
new network capacity.

This adaptation time can be split into two periods.  First, once the
slow-start phase has finished, the sender rate may not immediately start
increasing above $X_c$. Indeed, if there has not been enough packets sent during
the slow-start for $S_0$ to be larger than $S_1$ in~\eqref{eq:ImeanP}, $p$
will not decrease. During the loss recovery time, $X_\mathrm{Bps}$ is kept at
value $X_r^i=(R_\mathrm{old}/R_r^i) X_d$. After $t_\mathrm{recov}$, when enough
packets have been received, $p$ will start decreasing again. During this phase,
the sending rate slowly adapts to the available capacity, which is eventually
reached after $t_\mathrm{grow}$. In addition of
$n_\mathrm{wasted}$, a further $n_\mathrm{wasted}^\prime$ packets could be sent.
We develop a formulation of this extra capacity wastage below.

The loss recovery time, until the current loss interval contains enough packets
not to be ignored in~\eqref{eq:ImeanP}, is such that $0 < S_0 - S_1$ that is,
\begin{align}
  0 &< w_0 \underbrace{i_0}_{n_\mathrm{pkts}^{n_\mathrm{recov}}} +
  \sum_{n=1,\ldots,i-1} (w_n - w_{n-1}) i_n - w_{i-1} i_i.
  \label{eq:numerical-nrecov-imean-diff-generic}
\end{align}

To ``compete in the global Internet with TCP,'' it is recommended
in~\cite{rfc5348} to take $n=8$ and the weight vector as $\vec w = [1, 1, 1, 1,
0.8, 0.6, 0.4, 0.2]$. The formulation
of~\eqref{eq:numerical-nrecov-imean-diff-generic} can thus be simplified as
\begin{align}
  0 < n_\mathrm{pkts}^{n_\mathrm{recov}} - 0.2 \sum_{n=4}^8i_n \quad\rightarrow\quad &
  n_\mathrm{pkts}^{n_\mathrm{recov}} > 0.2 \sum_{n=4}^8i_n.
  \label{eq:numerical-npktsrecov-rfc5348}
\end{align}
When $n_\mathrm{pkts}^{n_\mathrm{recov}}$ packets have been sent since the reconnection,
$p$, driven by~\eqref{eq:ImeanP}, starts decreasing. It is difficult to estimate the
$i_n$ as they are dependent on the previous network conditions and specific
history. However, assuming a relatively stable network, all $i_n$ would be
similar and close to the inverse of $p_r$, the loss event rate of the previous
network. Thus, an estimate of the number of packets that need to be sent before
$p$ starts to adapt to the new network conditions can be written as
$n_\mathrm{pkts}^{n_\mathrm{recov}} = 1/p_r$.

This estimation allows to evaluate the duration of the recovery period,
$t_\mathrm{recov}$, which exists only if
$n_\mathrm{pkts}^{n_\mathrm{recov}}>n_\mathrm{pkts}^{n_\mathrm{ss}}$%
 (that is $t_\mathrm{recov}>0$) 
,
\begin{gather}
  t_\mathrm{recov} = \frac{s}{Xd}\left( n_\mathrm{pkts}^{n_\mathrm{recov}} - n_\mathrm{pkts}^{n_\mathrm{ss}} \right) = \frac{s}{Xd}\left( 1/p_r - n_\mathrm{pkts}^{n_\mathrm{ss}} \right).
  \label{eq:trecov}
\end{gather}

The additional amount of wasted capacity can be estimated as 
\begin{gather}
    \label{eq:numerical-rate-bigger}
    n_\mathrm{wasted}^\prime = \begin{multlined}[t]
      \frac1s (X_\mathrm{max} - X_d) \left( t_\mathrm{idle} + t_\mathrm{ss} + t_\mathrm{recov} \right) + \\
      \frac{R_\mathrm{new}}s \sum_{i=0}^{n_\mathrm{grow}} \left(X_\mathrm{max} - X_r^i \right)
    \end{multlined} \\
  \intertext{with}
  X_r^i = \begin{cases}
    X_d & \text{if $i=0$,}\\
    \min\left(X_\mathrm{Bps} \left( p_\mathrm{r} + \Delta p(n_\mathrm{pkts}^{i-1}, p_\mathrm{r}),R^i_r \right), 2X_r^{i-1} \right) & \text{otherwise.}
  \end{cases}
\end{gather}
Similarly to ~\eqref{eq:numerical-packets-wasted}, $n_\mathrm{grow}$ is the
number of RTTs needed to have $X^{n_\mathrm{grow}} \ge
X_\mathrm{max}$.

\subsection{Model Validation}

We now verify the model of the previous section by comparing numerical
results to \ns-2 simulations for a wide range of network parameters and
disconnection durations.

\begin{figure}[tb]
  \begin{center}
    \includegraphics[width=\columnwidth]{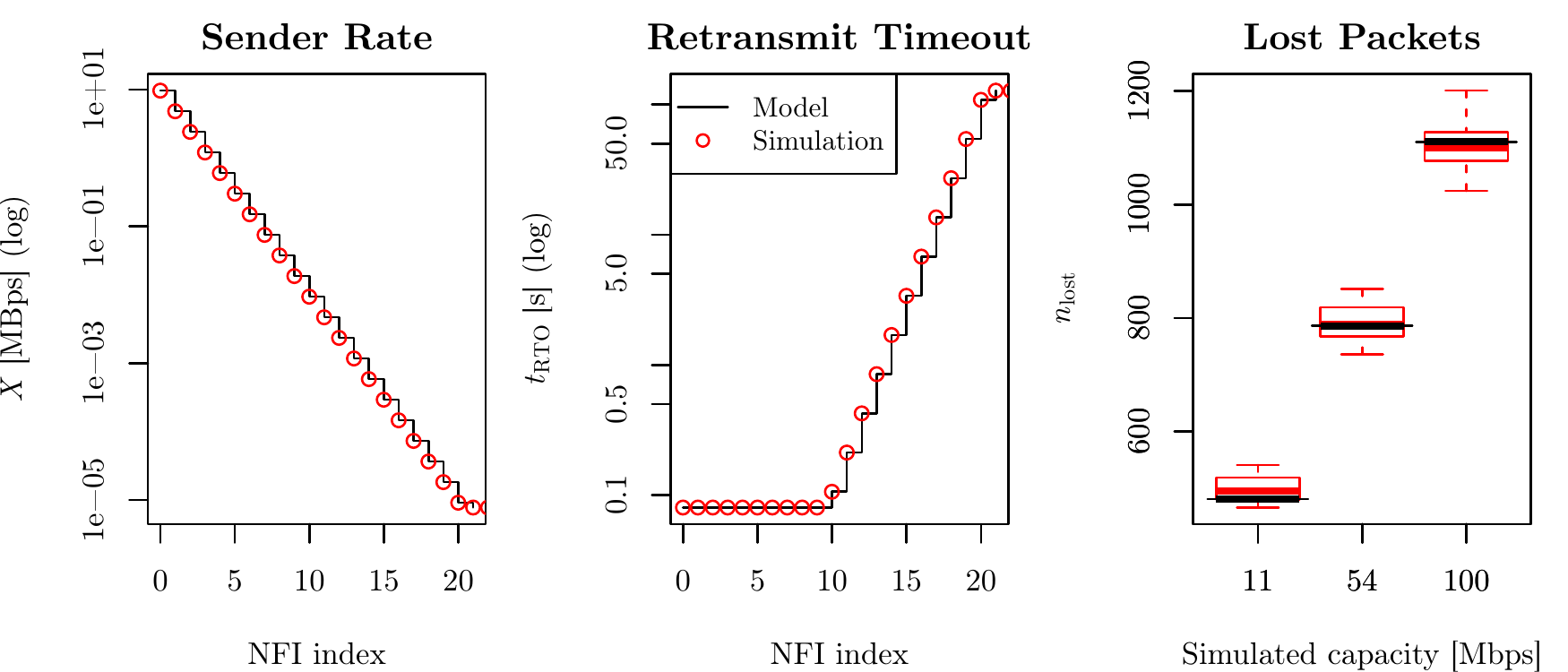}
  \end{center}

  \caption[TFRC model validation (internal parameters, lost packets and wasted
  capacity)]{Comparison of simulated TFRC's internal parameters, and number of
  lost packets during disconnections with the model's predictions. Results shown
  for $X=10$\,MBps and $R=2$\,ms.}

  \label{fig:model-check-disco-nlost-nwasted}
\end{figure}

To check the behaviour of TFRC during the disconnected
period---\eqref{eq:numerical-x-i} and \eqref{eq:numerical-trto-i}---as well as
the resulting number of lost packets~\eqref{eq:numerical-packet-losses}, 60\,s
disconnections are introduced after a variable amount of time, for link
capacities of 10, 54 and 100\,Mbps and for a range of delays (1--100\,ms).  The
number of packets sent after the disconnections is then counted in the
simulation trace file.  \figurename~\ref{fig:model-check-disco-nlost-nwasted}
shows a comparison of simulation results with predictions from the model for
$R=1$\,ms. It confirms that our model exactly predicts the values of the internal
parameters of the TFRC sender, and accurately estimates the number of losses.

The number of wasted packets, as determined
by~\eqref{eq:numerical-packets-wasted} and~\eqref{eq:numerical-rate-bigger}
cannot be derived in the same way from the \ns-2 trace files. Detecting the end
of the adaptation periods $t_\mathrm{ss}$ and $t_\mathrm{grow}$ relies on
comparing the current rate to $X_\mathrm{max}$, for which it is impossible to
obtain a ground truth from the simulations. It is therefore impossible to
identify over which period to count the additional packets which could have
potentially been sent, and the results vary depending on the estimate taken for
$X_\mathrm{max}$. In the following, we take $X_\mathrm{max} = X_\mathrm{recv}$
(from \tablename~\ref{tab:stationary-params} below); the order of magnitude of the
resulting numbers for the wasted capacity are coherent, but the reported value
is only indicative, and should not be used in further derivations.

Even though the presented model does not encompass all the details of the
behaviour of a real TFRC sender, it has proven to have sufficient prediction
accuracy to be used in estimating potential performance gains.

\subsection{Potential for Improvement}

We use our numerical model to determine the performance improvements that can be
expected from a better handling of disconnections. The input parameters ($X_d$
and $R$ of both links) for the model are those observed by TFRC when the
stationary state has been reached. These are summarised in
\tablename~\ref{tab:stationary-params} (based on network characteristics
presented later in~\tablename~\ref{tab:characteristics}). 

\begin{table}[tb]
  \centering
  \caption[Network parameters as observed by the \textit{ns}-2 TFRC sender]{Network
  parameters as observed by the \ns-2 TFRC sender in the stationary phase,
  reached at $T_\mathrm{stat}$.}
  \label{tab:stationary-params}
  \begin{tabular}{cccc}
    \toprule
    \textbf{Link type} & \textbf{$X_\mathrm{recv}$ [MBps]} & \textbf{$R$ [s]} & \textbf{$T_\mathrm{stat}$ [s]} \\
    \midrule
    UMTS & 	0.044 & 	0.96 & 	660.54 \\
    802.16 & 	1.10 & 	0.17 & 	264.14 \\
    802.11b & 	1.27 & 	0.05 & 	50.69 \\
    802.11g & 	4.82 & 	0.04 & 	21.67 \\
    \bottomrule
  \end{tabular}
\end{table}

MIPv6 relies on message exchanges over the network to update the binding
with the HA.\footnote{Fast handovers~\cite{2001koodli_fast_handover_context_transfer} could be used to reduce
the disconnection duration. However, packets arriving during the hand-off are
buffered at the new access router, which is not desirable for real-time traffic
as this would create latency at the application layer upon completion of the
handover.} Therefore, handover delays will vary depending on the characteristics of
the current link. The RTT on the new access network is particularly important
as it affects the delay until the MIPv6 binding updates are received.
In~\cite{2004lee_mipv6_handoff_analysis}, the authors thus proposed to use $t_\mathrm{ho} = 2.5
+ R$ as the time to complete the handover, and during which packets cannot be
successfully transmitted to the MN. We use this model as $t_D = t_\mathrm{ho}$.

An estimate of the stationary phase value for RTT on the new link is used for
$R$ in $t_\mathrm{ho}$. It is an over-estimation as the RTT is likely not to be
as high upon reconnection as when the full rate is established. Therefore, the
presented results should be considered an upper bound for the packet loss and
lower bound for the wasted capacity.

The number of lost packets and the available capacity wasted during a MIPv6
handover, as predicted by the model, are shown on
\tablename~\ref{tab:model-mip-nlost-nwasted}.  These results confirm that the
behaviour of TFRC can be improved. We intend to provide such improvement in
the next section.

\begin{table}
  \centering

  \caption[Losses and wasted capacity during a handover]{Predicted packet losses and
  wasted available capacity expected during a MIPv6 handover.}

  \label{tab:model-mip-nlost-nwasted}
  \begin{tabular}{ccccc}
    \toprule
    \multirow{2}{*}{\backslashbox{\textbf{from}}{\textbf{to}}} & \multirow{2}{*}{UMTS} & \multirow{2}{*}{802.16} & \multicolumn{2}{c}{802.11} \\
    & & & b & g \\
    \midrule
    \multicolumn{5}{c}{\textbf{Packet losses}} \\
    \midrule
    UMTS	& 306	& 236	& 226	& 224	\\
802.16	& 2760	& 2614	& 2614	& 2614	\\
802.11b	& 1080	& 1078	& 1078	& 1078	\\
802.11g	& 2909	& 2907	& 2907	& 2907	\\

    \midrule
    \multicolumn{5}{c}{\textbf{Wasted capacity [Number of 500\,B packets]}} \\
    \midrule
    UMTS	& 0	& $8\times10^4$	& $3\times10^2$	& $1\times10^5$	\\
802.16	& 0	& $5\times10^2$	& $2\times10^2$	& $1\times10^3$	\\
802.11b	& 0	& 0	& $1\times10^3$	& $5\times10^4$	\\
802.11g	& 0	& 0	& 0	& $5\times10^3$	\\

    \bottomrule
  \end{tabular}
\end{table}

\section{Freezing the DCCP/TFRC Transmission Upon Disconnections}
\label{freezetfrc:freezing}

In this section, we present an enhancement and its implementations to approach
the possible improvement made apparent above.  This modification relies on two
main additional stages. The sender's state is first frozen just before a
hand-off so as not to disrupt its performance, and transmission is suspended.
When the handover is complete, the sender is unfrozen. Then, with assistance
from the receiver it restores its previous rate and, if possible, probes the new
network path for a larger usable capacity.  Though this work primarily addresses
break-before-make handovers, a subset can also apply to make-before-break
events. In these cases, only the probing phase is needed.

The rationale for restoring the previous rate comes from
Freeze-TCP~\cite{2000goff_freezetcp}, where only wireless fading or horizontal
handovers were considered. Another advantage of this approach is that this rate
is known without dependence any external information source. In the case of
vertical hand-offs, as we consider them here, other approaches could also be
envisionned.  Some additional cross-layer interaction would allow the
application or some management element to specify another restart point based on
requirements or external knowledge of the expected
capacity~\cite[\eg,][]{1999balakrishnan_integrated_congestion_control,2011mehani_multihomed_flow_management}.
%
%
However, it would be important to ensure that the selected rate is not so high
as to cause overly excessive congestion on the new path during the first RTT.
In the following, we only consider previous-state restoration.

\subsection{Rationale of the Improvements}

Beyond the packet losses and under-usage of the available capacity, a reduction
in the immediate rate is quite detrimental to real-time applications. As
previously shown in Section~\ref{sec:after-reco}, it can take
up to several seconds to restore the rate after the completion of a handover.
During this period, applications observe high error rates as they cannot fit the
required amount of data units in the rate allowed by the transport protocol,
which results in bad QoE.  In such a situation, restarting from the rate
achieved before the handover would enable the application to drastically reduce
this period of bad quality.

Considering a communication involving video, the user's experience can be
further improved if the new access network has better characteristics, such as a
larger capacity. In the new network, the video codec could use a higher encoding
rate which would increase the overall QoE. Including a mechanism to probe the
new network path can make the new available path capacity available much faster
to the application. It could then take advantage of this larger capacity to
enhance the overall user experience.

Finally, having information about upcoming handovers, it is also possible to
limit, or even nullify, the number of lost packets. Doing so also has the
advantage of avoiding the unnecessary use of the old network path to send data
which would never be received as the receiver has moved.

When a hand-off is imminent (as detected through %
, \eg, IEEE 802.21~\cite{2009piri_80221} or 
control frameworks
%
%
~\cite{2011mehani_multihomed_flow_management}), we thus propose to
temporarily suspend the evolution of specific internal parameters. The sender
keeps an estimate of the current stationary parameters of the network used to
derive the sending rate offered to the application. Further packet transmission
is also prevented as the path from the sender to the receiver is known to be
temporarily cut.  When connectivity is available anew, the rate is restored
immediately and adapted as soon as possible to the new network conditions. The
congestion control algorithm first allows packets to be sent at the same rate as
before. If no error is reported, the sender then tries to probe the network path
by increasing its rate. In a way similar to the initial slow-start, the sending
rate doubles every RTT until the capacity of the new network has been reached.

\subsection{New States to Support the Freezing Mechanism}

We implement our proposed enhancement to TFRC within DCCP's CCID~3. The
operation of the resulting Freeze-DCCP/TFRC is separated into three phases:
\emph{Frozen}, \emph{Restoring} and \emph{Probing}. New states are implemented
into the TFRC sender and receiver to support these phases. Additionally, new
DCCP options are introduced to enable the required freeze/unfreeze signalling,
while state transition and synchronisation is done purely through TFRC options.

\begin{figure*}[tb]
  \centering
  \subfloat[Freeze-DCCP/TFRC sender]{\resizebox{.95\columnwidth}{!}{
  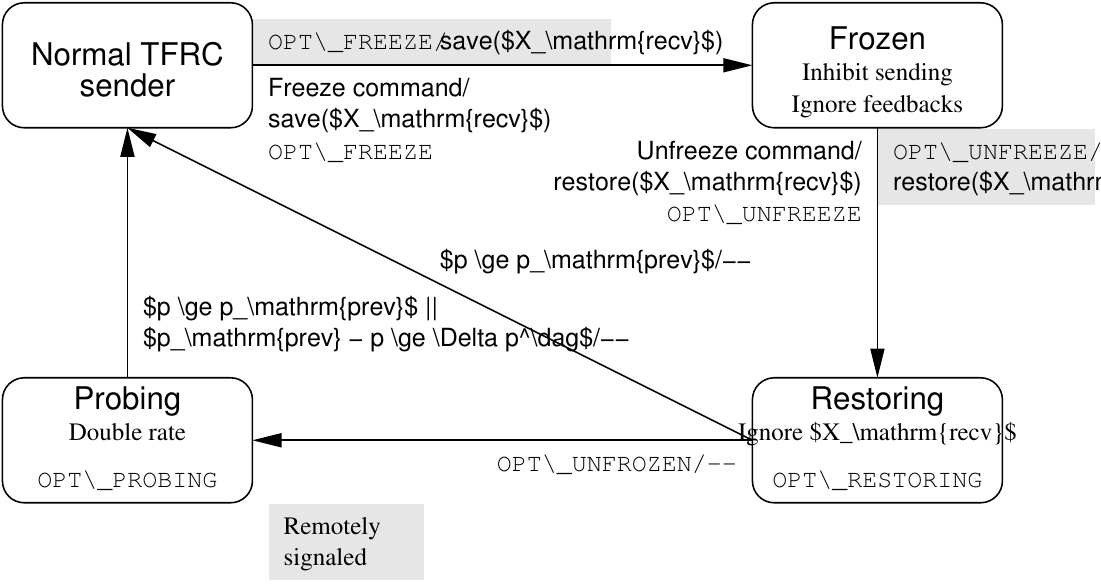
  \label{fig:freeze-states:sender}
  }}
  \hfill
  \subfloat[Freeze-DCCP/TFRC receiver]{\resizebox{.95\columnwidth}{!}{
  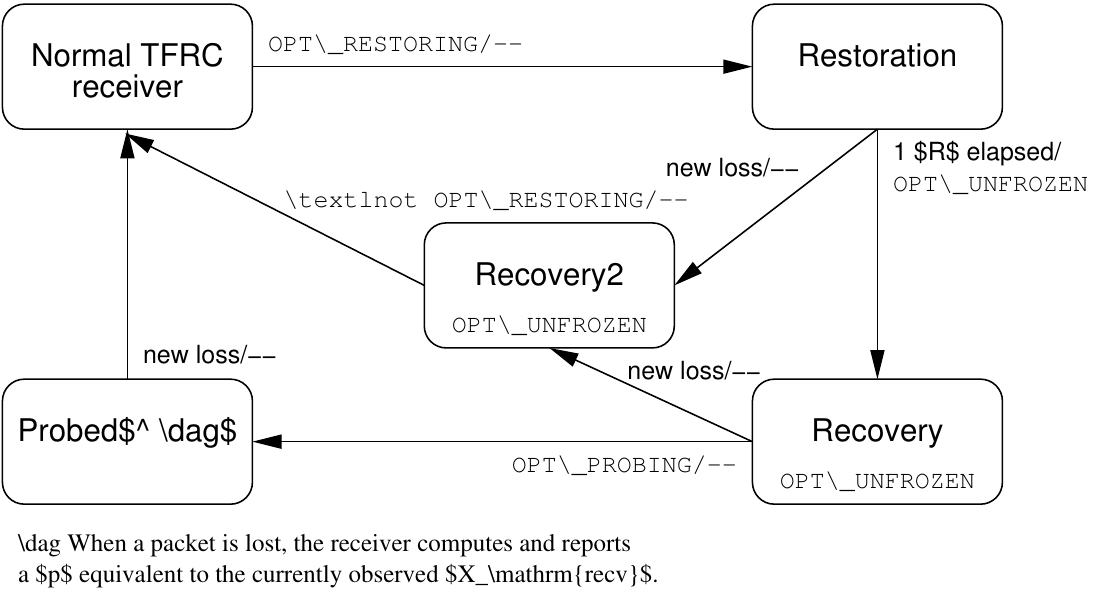
  \label{fig:freeze-states:receiver}
  }}

  \caption[Additional states and option exchanges to support
  Freeze-DCCP/TFRC]{Additional states and option exchanges to support
  Freeze-DCCP/TFRC (transitions are labelled as Condition/Action).  The
  sender~\subref{fig:freeze-states:sender} can be instructed to freeze or
  unfreeze either locally or by the remote peer. The
  receiver~\subref{fig:freeze-states:receiver} does not have to enter a Frozen
  state, but must perform some specific tasks during the Restoration and Probed
  phases.  Options can signal both/either the remote sender and/or receiver.}

  \label{fig:freeze-states}
\end{figure*}

\figurename~\ref{fig:freeze-states} shows the proposed Freeze-DCCP/TFRC state
diagram. The sender has three new states, shown in
\figurename~\ref{fig:freeze-states:sender}.  As most of Freeze-DCCP's operation
is driven by the sender, its states are directly named after the three phases.
The receiver has two ``active'' states: \emph{Restoration} and \emph{Probed}.
Both \emph{Recovery} states are transient and used to ensure synchronisation.
These are shown in \figurename~\ref{fig:freeze-states:receiver}.

The following sub-sections detail the signalling options and the evolution of the
states, as well as their specific actions throughout the Freeze-DCCP/TFRC phases.

\subsection{Additional Signalling}

Though window-based flow-control mechanisms have been proposed for
TFRC~\cite{2010lochin_chameleon}, they have not been included in the standard.
Thus, unlike Freeze-TCP~\cite{2000goff_freezetcp}, it is not possible to freeze
a DCCP/TFRC sender by simply reporting a specific value in a feedback message.
Additionally, it is desirable to be able to locally suspend the sender. To fully
support freezing on both sides, it is necessary to introduce new signalling
options, to be carried in the DCCP or TFRC packet headers. Our proposal does
not, however, change or extend the format of these headers; the options will be
gracefully ignored by standard implementations.

In a generic typical case, a mobile node, both sending and receiving over the
same DCCP connection, will detect or be informed that it is about to lose its
current connectivity. In this proposal, packets with an \verb#OPT_FREEZE#
DCCP-level option will be sent to the remote peer to suspend its sender
operation, then the local sender will be frozen. As the packets carrying the
option have to make it to the remote sender, the transport layer has to be
informed about upcoming disconnection a short while before it is to happen. A
minimum of one RTT is usually considered a reasonable
value~\cite{2000goff_freezetcp}. This delay is enough for signalling packets to
arrive on time (1/2 RTT later) to prevent the transmission of messages which
would have arrived after the disconnection (another 1/2 RTT later).

When connectivity becomes available again, the mobile node can restart its
traffic and instruct its peer to act similarly by sending packets with
an \verb#OPT_UNFREEZE# option.

Additional TFRC-level options are used to support further signalling during the
unfreezing phases. The sender uses \verb#OPT_PROBING# and \verb#OPT_RESTORING#
to indicate its current state, while the receiver sends an \verb#OPT_UNFROZEN#
to signal that it is ready for the Probing phase.

As DCCP is an unreliable protocol, option-carrying packets can be silently
lost. Extra care must be taken to ensure both peers are synchronised. This can
be done by exchanging options in a redundant manner. The naive approach of
adding those to every outgoing packet is chosen here. Depending on the
application, this however risks consuming too much capacity and reduction of the
option frequency could be considered. As for the \verb#OPT_FREEZE# and
\verb#OPT_UNFREEZE# options, an implementation should take care of repeating
them on several packets (three in our implementations).

\subsection{Frozen Phase}

When instructed to freeze, either locally or by the remote peer, the sender
enters the Frozen state. In this state, all data transmission ceases. This
ensures that no packet will be lost. It in turn guarantees that the loss event
rate calculated by the receiver will be kept unmodified. The receiver does not
need any specific state for this phase.

The disconnection may however not happen right after freezing, and additional
feedback from the receiver may arrive at the sender. Parameters such as the
RTT $R$, or the receiver rate $X_\mathrm{recv}$ risk being updated. Thus,
while in the Frozen state, the sender ignores all feedback messages. When
entering this state, it also saves the value of $X_\mathrm{recv}$ as it will be
locally modified on every expiration of the \verb#nofeedback# timer.

To efficiently address longer disconnection periods, and provide some sort of
DTN support, it is advisable to additionally increase the connection timeout.
Indeed, disconnections longer than 8 minutes may result in the frozen socket
being prematurely closed.\footnote{According to \cite{rfc4340}, a socket in the
\emph{Respond} state waits a maximum of four Maximum (TCP) Segment Life for
packets before resetting the connection. The Linux 2.6 implementation
generalises this idle timeout to the entire lifespan of the socket.}

\subsection{Restoring Phase}

After receiving a local unfreeze instruction or the \verb#OPT_UNFREEZE# option,
the sender enters the Restoring state. It first restores $X_\mathrm{recv}$.  The
send timer is then reset to resume packet transmission.  As the parameters are
the same as before the disconnection, the sending rate will be restored to its
previous value.

At the same time, it is no longer necessary to completely ignore feedback from
the receiver. It is however needed to keep ignoring the $X_\mathrm{recv}$
reports. Indeed, the receiver rate is measured over at least one RTT. The first
feedback packets are likely to cover part of the disconnected period resulting
in an incorrectly low value for $X_\mathrm{recv}$. Using such value may create
instabilities in the sending rate as it is bound by $2X_\mathrm{recv}$ as
per~\eqref{eq:tfrc-update-x}.

When in the Restoring state, the sender adds an \verb#OPT_RESTORING# option to
all its outgoing packets to put the receiver into the Restoration state.  The
Restoring phase ends when the loss event rate increases or an
\verb#OPT_UNFROZEN# option is received. This option is added by the receiver
after a complete RTT has elapsed, thus signalling that it is no longer
necessary to ignore the value of $X_\mathrm{recv}$ as it will now correctly
reflect the receiver rate.

\subsection{Probing Phase} 

Standard TFRC quickly reacts to a reduction in the available capacity by
responding promptly to an increase in the loss event rate. The
\verb!conservative_! mode outlined by
\cite{2001bansal_slowly_responsive_congestion_control} further increases this
response. Conversely, after idle periods, it is proposed
in~\cite{draft-ietf-dccp-tfrc-faster-restart-06} to increase the sending rate
back to the previously supported maximum at an increased pace by quadrupling the
rate every RTT.\footnote{This draft proposal considers application-limited rates
rather than disconnections; it also does not restore the rate at once as our
Restoring phase does.} There is however no mechanism to quickly adapt to
\emph{better} network conditions. In the Probing state, our sender checks for
such improvement in the new network. This is done only if no loss has occurred
during the Restoring phase. The sender uses the \verb#OPT_PROBING# option to
inform the receiver of its new state. Upon receiving this option, the receiver
enters the Probed state.

This phase is similar to a slow-start. Every RTT, the sending rate is
doubled.  When a loss is detected while it is in the Probed state, the receiver
reinitialises its loss history to match the last measured rate. It first
computes a packet loss rate $p$ equivalent to the observed receiver rate
$X_\mathrm{recv}$. It then reinitialises a complete history of $n$ loss
intervals of the calculated size.

As $p$ is completely recomputed by the receiver on the first loss, it can be
larger, lower or even equal to its previous value. The exit criterion for the
probing phase is therefore based on the expected evolution of the reported loss
event rate, as derived as~\eqref{eq:numerical-p-variation} in
Section~\ref{sec:model}. In a loss-less period, $p$ will never increase. With a
growing loss interval, it will however keep decreasing slightly. The sender
should thus exit the Probing state if
  $\Delta p \notin \left] \Delta p_\mathrm{min}(X_\mathrm{Bps} \cdot R, p_\mathrm{prev}); 0 \right[$, 
following~\eqref{eq:numerical-delta-p}. Missing \verb#OPT_PROBING#
options on new packets then takes the receiver out of the Probed state.

It may happen that the sender-recomputed $p$ lies in the acceptable range of
variation. In this case, the sender cannot detect that the Probing phase should
be ended. Some more losses will however be generated during the next RTT. These
losses will prevent $p$ from changing during the next report, thus properly
ending the Probing phase as per the previous criterion.

\section{Performance Evaluation}
\label{freezetfrc:performance}

This section  presents an evaluation of the enhancements proposed in the
previous section. It first compares, in \ns-2 simulations, the behaviour of
Freeze-DCCP/TFRC with that of the unmodified version. It then demonstrates that
the proposed mechanism still retains a satisfying level of fairness to TCP
flows.  Finally, a real experiment, based on a Linux implementation, shows that
our proposal is well suited to improve the QoE of a live video stream
experiencing multiple handovers between heterogeneous technologies.

\subsection{Realistic Handover Simulations}
\label{realistic}

Simulations were run with \ns-2.33. Additional modifications have been made to
the TFRC sender of the DCCP module%
~\cite{2004mattsson_dccp_ns2} 
to implement
the clarified loss average calculation of~\cite{rfc5348}. The
\verb#DCCP/TFRC/Freeze# agent has been implemented by deriving the DCCP/TFRC
C++ class to add the freezing mechanisms described above.

As the impact of disconnections is only relevant to our study at the transport
layer, the underlying wireless technologies are not simulated as such. Rather,
simple wired topologies are used, as suggested by
\cite{2004gurtov_modelling_wireless}. All wireless links are modelled as duplex
links, even the wireless ones. This may not be correct for bi-directional data
scenarios. Such scenarios, however, are not considered here. The
characteristics of the various technologies used in these simulation are taken
from the respective standards and measurement-based literature, and summarised in
\tablename~\ref{tab:characteristics}.

\begin{table}
  \caption{Characteristics of the wireless networks used for
  evaluation purposes.}
  \label{tab:characteristics}
  \centering
  \begin{tabular}{ccc}
    \toprule
    \textbf{Technology} & \textbf{D/L capacity [bps]} & \textbf{Avg. RTT [s]} \\
    \midrule
    UMTS & 384\,k & 250\,m~\cite{2005vacirca_umts_grprs_rtt_measurements} \\
    802.11b/g & 11\,M/54\,M & 20\,m~\cite{2005karapantelakis_ad-hoc_delay_experiments} \\
    802.16 & 9.5\,M~\cite{2007grondalen_wimax_measures} & 80\,m~\cite{2008halepovic_tcp_wimax} \\
    \bottomrule
  \end{tabular}
\end{table}

A router is placed between the sender and the receiver. Such a topology allows
to transparently disconnect the link between the router and the receiver without
preventing the sender from trying to transmit packets during the disconnections.
\verb#DropTail# queues with the default buffer size (50 packets) are used.

Disconnections are simulated by manipulating the routing model of \ns-2 with the
\verb#$ns_ rtmodel-at# function. The time of the hand-offs is chosen, once the
system is in a stationary state, from a uniformly distributed variable over a
time period of four RTTs. The generic behaviour of both standard TFRC and our
variant is thus captured. Link characteristics are modified using the
\verb#$ns_ bandwidth#
and \verb#$ns_ delay# commands.  The simulations were ended after the rates had
settled on the new network. The result have been averaged over 20
runs.

The Freeze-DCCP agent is instructed to suspend its connection locally (\ie, not
using the \verb#OPT_FREEZE# and \verb#OPT_UNFREEZE# options).  The \verb#freeze#
command is given one RTT  before the disconnection is scheduled to happen (as
suggested by~\cite{2000goff_freezetcp}). The \verb#unfreeze# instruction is
given 0.1\,ms after the network link is reconnected.

\begin{figure}[tb]
  \begin{center}
    \includegraphics[width=\columnwidth]{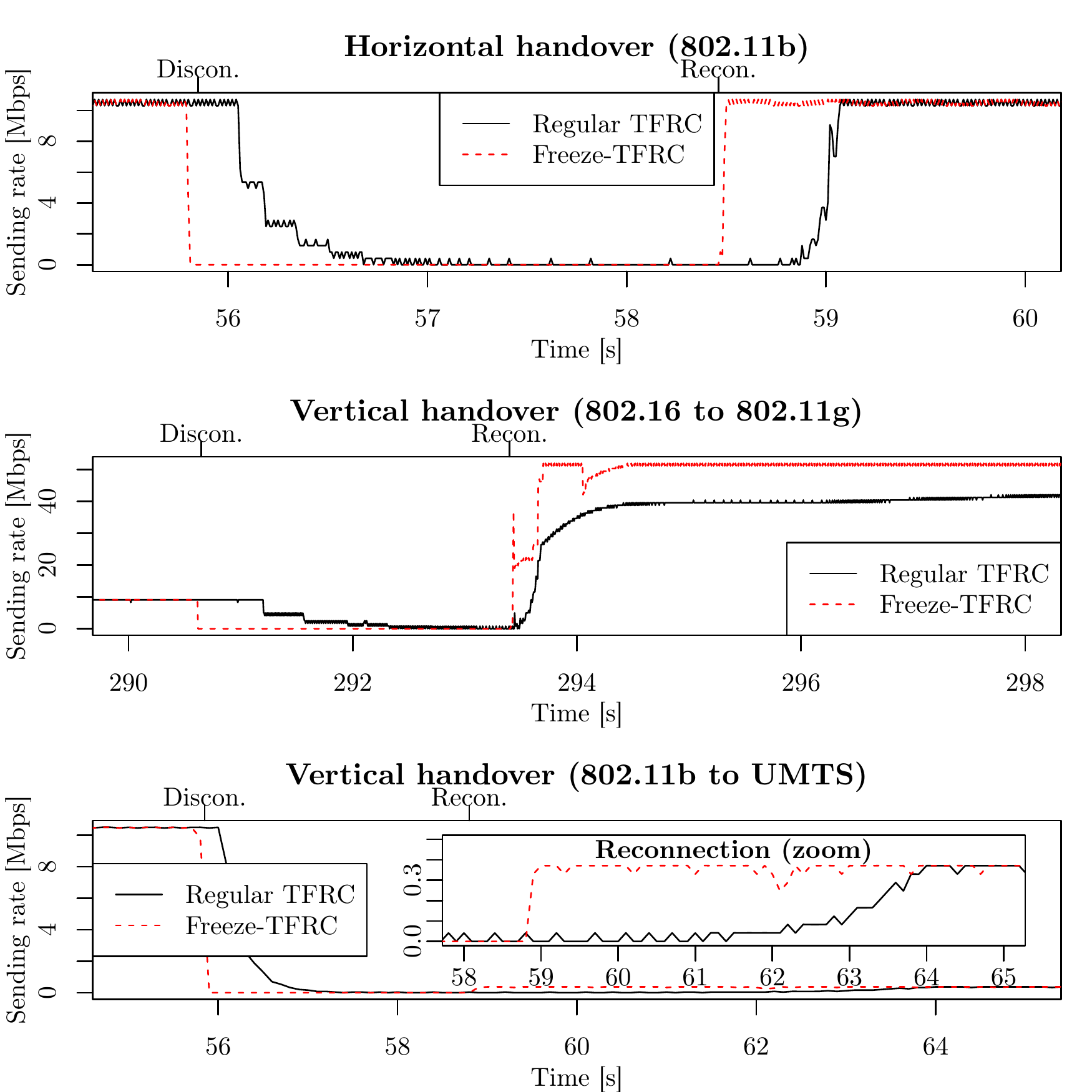}
  \end{center}

  \caption[Comparison of the rate of DCCP/TFRC\ and the Freeze-enabled version
  in typical examples of MIPv6 horizontal or vertical handovers]{Comparison of
  the rate of DCCP/TFRC and the Freeze-enabled version in typical examples
  of MIPv6 horizontal or vertical handovers.}
  
  \label{fig:sim-realistic}
\end{figure}

The number of losses upon reconnection, as well as the wasted capacity, are shown
on \tablename~\ref{tab:sim-mip-nlost-nwasted}. As Freeze-DCCP/TFRC did not
lose any packet, this information is omitted.  The wasted capacity has
been estimated by comparing TFRC's actual rate $X$ to what is achievable in the
steady state (\tablename~\ref{tab:stationary-params}), then converted in number
of 500\,B packets. \figurename~\ref{fig:sim-realistic} illustrates these
figures by comparing how both regular DCCP and the Freeze-enabled version
perform in key example scenarios.

\begin{table}[tb]
  \centering

  \caption[Comparison of DCCP/TFRC and Freeze-DCCP/TFRC in simulated MIPv6
  handovers]{Simulated MIPv6 handovers performance impact for DCCP/TFRC
  and Freeze-DCCP/TFRC (grey).}

  \addtolength{\tabcolsep}{-.5ex}
  \label{tab:sim-mip-nlost-nwasted}
  \begin{tabular}{ccccc}
    \toprule
    \multirow{2}{*}{\backslashbox{\textbf{from}}{\textbf{to}}} & \multirow{2}{*}{UMTS} &
    \multirow{2}{*}{802.16} & \multicolumn{2}{c}{802.11} \\
    & & & b & g \\
    \midrule
    \multicolumn{5}{c}{\textbf{Packet losses (DCCP/TFRC only)}} \\
    \midrule
    UMTS	& 253.3	& 269.8	& 273.6	& 275.4	\\
802.16	& 1732.3	& 1734.6	& 1734.6	& 1734.6	\\
802.11b	& 856	& 855.5	& 855.3	& 855.3	\\
802.11g	& 2470.9	& 2470.4	& 2470.2	& 2470.1	\\

    \midrule
    \multicolumn{5}{c}{\textbf{Wasted capacity [Number of 500\,B packets]}} \\
    \midrule
    UMTS & 50.5	& 54018.05	& 2190.45	& 92156.1	\\
\rowcolor{gray!20} --- & 13.4	& 3607.9	& 9342.75	& 89328.6	\\
802.16 & 12.45	& 1827.95	& 603.05	& 4185.75	\\
\rowcolor{gray!20} --- & 5	& 591.15	& 150.9	& 1520.35	\\
802.11b & 150.45	& 28314	& 2101.75	& 57970.65	\\
\rowcolor{gray!20} --- & 0	& 15278	& 47.45	& 1045.05	\\
802.11g & 42.5	& 2104.3	& 943.4	& 4313	\\
\rowcolor{gray!20} --- & 0	& 7172.75	& 46.5	& 188.45	\\

    \bottomrule
  \end{tabular}
\end{table}

Two cases stand out in \tablename~\ref{tab:sim-mip-nlost-nwasted} where the
average amount of wasted capacity is larger with our proposal than with standard
TFRC. These correspond to pathological cases for our \ns-2 implementation where
either the Recovery or the Probing phase experiences losses too early. The rate
reported by the receiver does not cover a full RTT (802.11 g to 802.16), or the
probing phase terminates before having discovered the network capacity (UMTS to
802.11 b). Future work should address this type of situation by using other or
additional metrics, rather than just losses, to drive the Restoration/Probing
phases.

Overall, Freeze-DCCP/TFRC quickly restores an acceptable rate for the
application, and greatly reduces the overall under-usage of the available
capacity upon reconnection. It also prevents traffic after the hand-off to
unnecessarily use the rest of the network.

\subsection{Fairness to TCP Flows}

TFRC was designed to quickly respond to reductions of the capacity. The
restoring and probing features of Freeze-DCCP/TFRC, however, aggressively test
and use the network. TCP-friendliness is however a key property of TFRC, which
it is desirable to retain. It is therefore important to check that our additions
do not make the protocol too greedy. Though~\cite{2007briscoe_fairness} argues
that usage fairness should not be based on rate comparisons, such an approach is
still commonly accepted, and we use it here.

The criterion to evaluate TCP-fairness is the ratio of the average capacity
occupation of Freeze-DCCP/TFRC to that of concurrent TCP flows. The samples, taken
after the reconnection, have been averaged over 100\,s, discarding the initial
rate settlement period.

\tablename~\ref{tab:tcpfairness} compares the average fairness of a
\mbox{(Freeze-)DCCP/TFRC} flow to a concurrent TCP stream, as observed after the
reconnection for the studied handover scenarios. The proposed improvement appears
to be reasonably fair to TCP flows in various scenarios including vertical
handovers to technologies with higher or lower capacities. In some cases it is
even too fair, not competing aggressively enough for the network. A similar
behaviour is however also observed for the regular DCCP/TFRC, and is not a
result of our changes.

\begin{table}
  \centering

  \caption[Fairness comparison of (Freeze-)DCCP/TFRC to TCP after a
  handover]{Fairness comparison of both standard TFRC and the Freeze-enabled
  proposal to TCP after a handover (taking network parameters from
  \tablename~\ref{tab:characteristics}). Values in the range $[0.5;2]$ are
  considered ``reasonably fair''~\cite{rfc5348}.}

  \label{tab:tcpfairness}
  \begin{tabular}{ccccc}
    \toprule
    \multirow{2}{*}{\backslashbox{\textbf{from}}{\textbf{to}}} & \multirow{2}{*}{UMTS} & \multirow{2}{*}{802.16} & \multicolumn{2}{c}{802.11} \\
    & & & b & g \\
    \midrule
    \multicolumn{5}{c}{\textbf{Standard DCCP/TFRC}} \\
    \cmidrule{1-5}
    UMTS	& 1.3	& 0.7	& 0.4	& 0.3	\\
802.16	& 1.1	& 1	& 0.9	& 0.8	\\
802.11b	& 1.2	& 1	& 1	& 0.8	\\
802.11g	& 1.3	& 1.1	& 1	& 1.1	\\

    \midrule
    \multicolumn{5}{c}{\textbf{Freeze-DCCP/TFRC}} \\
    \cmidrule{1-5}
    UMTS	& 0.6	& 0.3	& 0.2	& 0.1	\\
802.16	& 1.6	& 1.3	& 1.1	& 0.9	\\
802.11b	& 1.3	& 1	& 0.9	& 0.7	\\
802.11g	& 1.5	& 1.2	& 1	& 1.1	\\

    \bottomrule
  \end{tabular}
\end{table}

\subsection{QoE of Mobile Video Streaming}

To explore the actual performance improvements of our proposal, we used an OMF
testbed~\cite{2010rakotoarivelo_omf} to emulate vertical handovers and observe
the impact on the quality of a video stream.  An implementation of
Freeze-DCCP/TFRC was developed in the Linux kernel\footnote{%
We used the Net:DCCP tree 
(\url{http://eden-feed.erg.abdn.ac.uk/cgi-bin/gitweb.cgi?p=dccp_exp.git;a=summary}),
forked off of vanilla version 2.6.34-rc5, as a starting point. 
A Git branch containing these modifications is available at
\url{http://github.com/shtrom/linux-2.6/tree/freezedccp}.} and used here.

\begin{figure}[tb]
  \centering

  \resizebox{.75\columnwidth}{!}{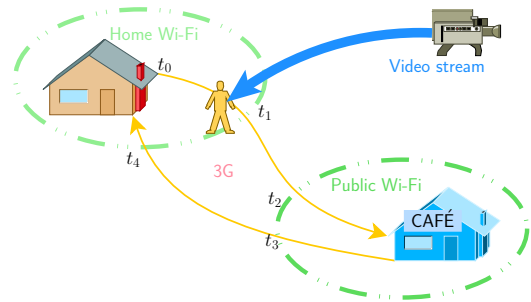}

  \caption[Scenario for the evaluation of Freeze-DCCP/TFRC improvements on
  application quality]{Scenario for the evaluation of Freeze-DCCP/TFRC
  improvements on application quality. A user viewing a video stream on their
  mobile device goes out for a coffee then comes back home. In the process,
  networks with heterogeneous characteristics are visited.}

  \label{fig:freezetfrc_psnr_scenario}
\end{figure}

The scenario is depicted in
\figurename~\ref{fig:freezetfrc_psnr_scenario}.\footnote{The OMF scripts are
available at
\url{http://omf.mytestbed.net/projects/omf-case-studies/wiki/FreezeDccpQoE}.} A
user, initially at home ($t_0$), receives a video stream on their mobile
terminal connected to their home Wi-Fi network connected to their DSL link
(1\,Mbps).  They decide to get a coffee from the corner shop. On the way there,
the mobile terminal loses its connectivity to the home network, and hands off to
a shared 3G network ($t_1$; 500\,kbps). The coffee shop has a public wireless network
(\eg, a shared Internet access, offering 700\,kbps of capacity), to which the
device connects when it arrives in range ($t_2$).  With their coffee in hand,
the user then heads back home, losing connectivity to the public Wi-Fi network
and performing a new handover to 3G at $t_3$ before finally reconnecting to
their home network at $t_4$.  Throughout the streaming period, the device thus
goes through several handovers between various wireless networks with different
capacities and delays.

In this experiment, a video file, encoded and packetised using the H.264 codec
with a 1\,Mbps bit-rate, is sent using a specially patched version of
Iperf%
~\cite{2004gates_iperf2}
\footnote{This version, supporting both DCCP and
OML, is available from
\url{http://www.nicta.com.au/people/mehanio/freezedccp\#iperf-dccp-oml}.} to a
custom receiver application. Both endpoints provide information about the
sending or receiving rates. The custom receiver identifies lost packets and
computes a moving average of the PSNR over 24 frames ($\simeq1$\,s), 
%
%
%
using the \sysfun{compare} tool of the ImageMagick
package~\cite{2005still_imagemagick}.\footnote{This tool computes the PSNR as
the $\log$ of the mean square error between an image and its reference.}
%
%
Both sender and receiver have been instrumented with the OML
toolkit~\cite{2013mehani_instrumentation_framework}. This allows them to report
readings of these metrics in real time for analysis or display.

Our OMF testbed currently only supports 802.11-based wireless networks.
However, only a limited set of parameters of the underlying network is relevant
at the transport layer. Therefore, as for the previous section, we follow the
suggestion of~\cite{2004gurtov_modelling_wireless} and emulate the conditions of
wireless technologies in different networks by shaping the available
capacity using Linux' traffic control tools~\cite{2004lartc_howto},  and
regulating forwarding delays using NetEm%
~\cite{2005hemminger_netem}
.

\begin{figure}[tb]
  \centering
  \includegraphics[width=.75\columnwidth]{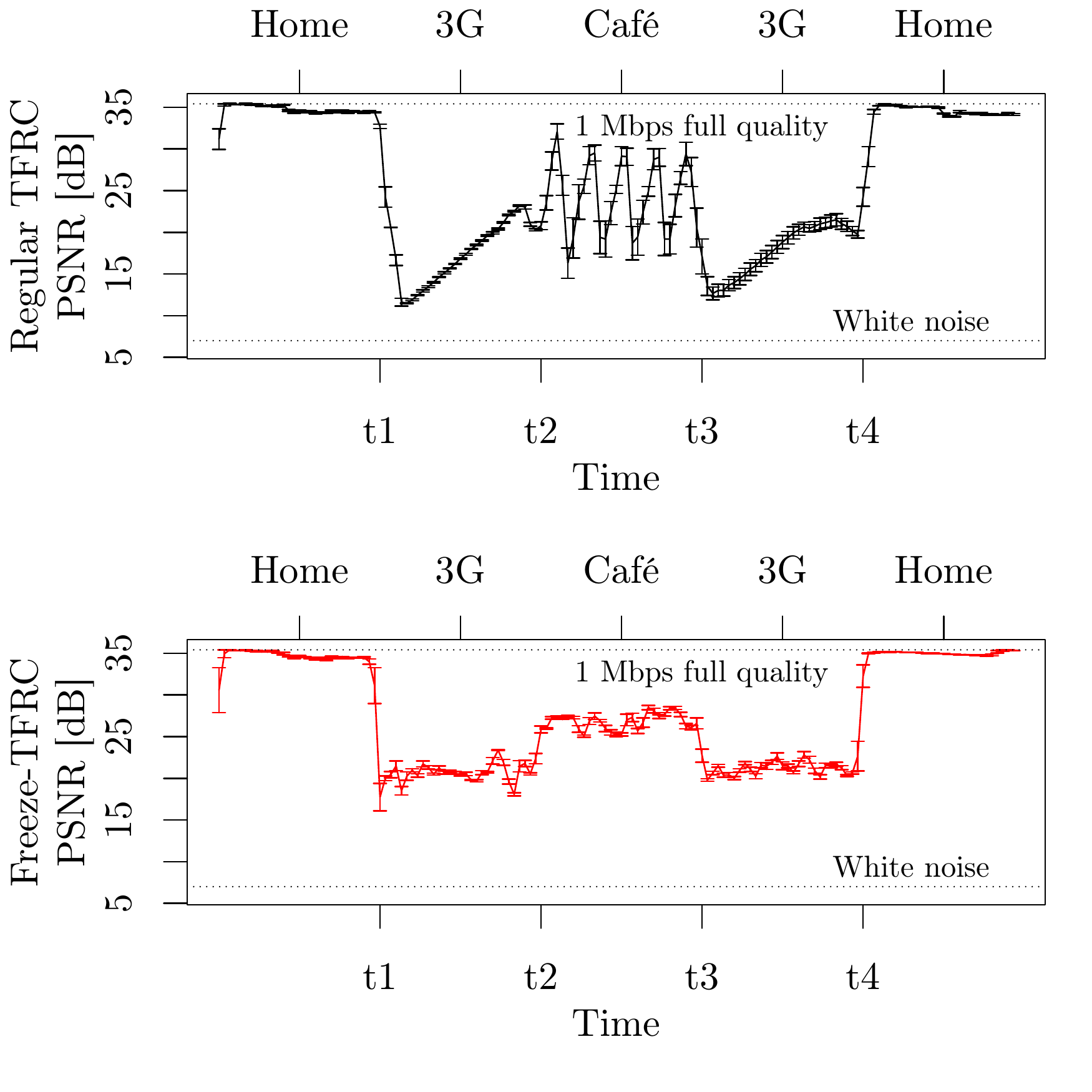}
  \caption[PSNR comparison for a video stream using TFRC or Freeze-TFRC in
  mobility situations]{PSNR comparison for a video stream using TFRC or
  Freeze-TFRC in mobility situations. Averages over 8 runs; error bars
  indicate the standard error.}
  \label{fig:freezetfrc_psnr}
\end{figure}

The results, averaged over 8 runs of our scenario, comparing the PSNR of the
video stream when using Freeze-DCCP/TFRC to that with the standard version, are
shown in \figurename~\ref{fig:freezetfrc_psnr}. They show that the use of
Freeze-DCCP/TFRC results in a reduced but stable QoE when on those networks
which cannot support 1\,Mbps streams, as our proposal is able to adapt much
faster and use close to the full available path capacity to carry application
data. In comparison, the PSNR of a video stream supported by the regular
DCCP/TFRC reduces to a minimum (a PSNR of 7\,dB is that of a purely random
image), and takes a long time (up to the complete visit duration of a network)
to restore to a better level. In addition, Freeze-DCCP/TFRC does not suffer from
the oscillations which appear for the standard version when visiting the Café's
Wi-Fi network. We hypothesise that it is due to the probing mechanism finding
the capacity of the new network more accurately, and its estimates not being
biased by measurements from the previous network. It might be interesting to
instrument the kernel code of DCCP (\eg, TFRC's loss history) in order to
investigate this further.

\section{Conclusion and Future Work}
\label{freezetfrc:conclusion}

In this article, we have first identified the issues that TFRC faces in
mobility situations.  We have numerically modelled the losses and subsequent
under-usage of the available capacity that it experiences in those cases. This
model allowed us to evaluate the performance improvements that could be expected
from a system with a better awareness and handling of disconnections.  We thus
proposed Freeze-DCCP/TFRC, an extension of the TFRC congestion control
mechanism used by DCCP, to approach these possible performance gains. This
proposal is aimed at uses of DCCP in situations where network connectivity may
periodically not be available for varying periods of time, and the access
networks' characteristics may widely vary between disconnections.

Freeze-DCCP/TFRC was both implemented in \ns-2 and Linux. Simulation results
have shown that it is possible to prevent handover-induced losses, to restore
the rate faster when reconnecting to a link with lower or similar capacities and
to adapt more quickly to higher capacities.  Additionally, we confirmed that our
proposal maintains the important TFRC's property of being fair to concurrent TCP
flows.  We have also experimentally shown that our proposal can significantly
improve the performance quality of streaming applications when disconnections
are predictable, for example, for IP mobility or.  Though the proposed
modifications were designed with real-time applications over DCCP in mind, we
believe they are versatile enough in terms of rate adaptation and packet loss
avoidance to also benefit other types of traffic.

Additional work may however be needed for the proposed extension to be used in
real deployments.  First, it would be desirable to decouple the freezing
mechanism, which caters for hand-off-induced disconnections, and the probing
phase, which deals with heterogeneous paths. Indeed, only the latter is needed
for make-before-break handovers. Second, the current probing mechanism assumes
packets are lost only when the capacity of the new network is reached.  Other
methods of detecting that the current rate matches the available path capacity
should be explored.  Both in-band solutions, such as MBTFRC, and out-of-band
ones could be considered.

\printbibliography 

\end{document}